\documentclass[aps,pre,twocolumn,superscriptaddress,showkeys,showpacs,longbibliography,groupedaddress,floatfix,noeprint]{revtex4-2}
\pdfoutput=1

\usepackage[utf8]{inputenc}
\usepackage{graphicx}
\usepackage{amssymb, amsmath}
\usepackage{color,soul}
\usepackage{array}
\usepackage{float}

\begin{document}
 
\title{Zero-Temperature Coarsening in the Two-Dimensional Long-Range Ising Model}
  
\author{Henrik Christiansen}
\email{henrik.christiansen@itp.uni-leipzig.de}
\author{Suman Majumder}
\email{suman.majumder@itp.uni-leipzig.de}
\author{Wolfhard Janke}
\email{wolfhard.janke@itp.uni-leipzig.de}
\affiliation{Institut für Theoretische Physik, Universität Leipzig, IPF 231101, 04081 Leipzig, Germany}
\date{\today}

\begin{abstract}
We investigate the nonequilibrium dynamics following a quench to zero temperature of the non-conserved Ising model with power-law decaying long-range interactions $\propto 1/r^{d+\sigma}$ in $d=2$ spatial dimensions.
The zero-temperature coarsening is always of special interest among nonequilibrium processes, because often peculiar behavior is observed.
We provide estimates of the nonequilibrium exponents, viz., the growth exponent $\alpha$, the persistence exponent $\theta$, and the fractal dimension $d_f$.
It is found that the growth exponent $\alpha\approx 3/4$ is independent of $\sigma$ and different from $\alpha=1/2$ as expected for nearest-neighbor models.
In the large $\sigma$ regime of the tunable interactions only the fractal dimension $d_f$ of the nearest-neighbor Ising model is recovered, while the other exponents differ significantly.
For the persistence exponent $\theta$ this is a direct consequence of the different growth exponents $\alpha$ as can be understood from the relation $d-d_f=\theta/\alpha$; they just differ by the ratio of the growth exponents $\approx 3/2$. 
This relation has been proposed for annihilation processes and later numerically tested for the $d=2$ nearest-neighbor Ising model.
We confirm this relation for all $\sigma$ studied, reinforcing its general validity.
\end{abstract}

\maketitle
\section{Introduction}
The nonequilibrium dynamics of systems quenched from a random start configuration to an ordered state is of fundamental interest and has been studied in numerous systems, ranging, e.g., from spin systems \cite{bray2002theory,puri2009kinetics,krapivsky2010kinetic} to polymers \cite{majumder2015cluster,christiansen2017JCP,majumder2017SM,majumder2019understanding}.
The process of coarsening or phase ordering kinetics of a system starting from a random starting configuration to zero temperature, i.e., during a pure energy-minimization procedure, can be described and {\em classified} by a number of nonequilibrium exponents.
One can directly observe the growth of ordered regions in every such process, which is quantified by estimating the characteristic length $\ell(t)$ and the associated growth exponent $\alpha$.
Another class of observations can be summarized under the term ``persistence'', which is a concept directly related to the first-passage properties of systems \cite{redner2001guide}.
It is defined only for quenches to $T \equiv 0$, although there have been attempts to also extract these properties from simulations at finite $T$ \cite{derrida1997extract,cueille1997spin}.
We investigate both those properties and related observables for the long-range Ising model with algebraically decaying interactions, for which very little is known both analytically and numerically.
This process is especially interesting at zero temperature, since exceptional behavior has been shown to often be observable under such special circumstances.
\par
In the following Section \ref{ScalingPred} we first recall the models and present some scaling predictions, followed by a discussion of the used methods in Section~\ref{Methods}.
Subsequently results are presented which are expected to be in the short-range-like regime of the long-range Ising model \cite{Bray_only,bray1994growth,Rutenberg1994,christiansen2018,ConferenceChristiansen,corberi2019one,christiansen2019non} in Section~\ref{SR}.
Next in Section \ref{LR}, we will have a closer look at the ``truly'' long-range regime which are still treatable without too overwhelming finite-size effects.
The behavior for intermediate interaction strengths will be investigated in Section \ref{diffSig} before concluding in Section \ref{conclusion}.

\section{Models and Scaling Predictions}
\label{ScalingPred}

The most studied model in coarsening phenomena is the nearest-neighbor Ising model (NNIM) with Hamiltonian
\begin{equation}
\mathcal{H}=-J \sum_{\langle ij \rangle} s_is_j,
\end{equation}
in $d=2$ spatial dimensions on a square lattice with periodic boundary conditions.
In the Hamiltonian, $s_i=\pm 1$ are the spins and $J$ is the coupling constant, where $J > 0$ for ferromagnetic interactions.
Here $\langle ij \rangle$ symbolizes a summation over all nearest-neighbor pairs.
The main system under consideration is an alteration, where every spin interacts with all other spins, the long-range Ising model (LRIM) with Hamiltonian
\begin{equation}
  \mathcal{H}=-\sum_{i<j}J(r_{ij})s_is_j.
  \label{HLRIM}
\end{equation}
Here $J(r_{ij})$ is the power-law decaying potential of form 
\begin{equation}
  \label{Hamil}
  J(r_{ij})=\frac{1}{r^{d+\sigma}},
\end{equation}
where $s_i=\pm 1$ are again spins located on a square lattice with periodic boundary conditions in $d=2$ spatial dimensions.
\par
In most cases, the growth of the characteristic length is given by a power law 
\begin{equation}
  \label{len}
  \ell(t) \sim t^{\alpha},
\end{equation}
where $\alpha$ is the growth exponent.
For many models, such as the NNIM in $d=2$, it is well established that the nonequilibrium growth exponent is directly related to the equilibrium dynamical exponent, i.e., $\alpha = 1/z$, where with non-conserved order parameter $z=2$ for all quench temperatures $T<T_c$.
\par
Persistence \cite{majumdar1999persistence,bray2013persistence} has been studied in various other contexts, ranging from random walk like systems \cite{majumdar1996nontrivial,majumdar1996survival,bauer1999statistics,sire2000analytical,ehrhardt2004persistence} over surface growth \cite{krug1997persistence,kallabis1999persistence,constantin2003infinite,constantin2004persistence} to disordered systems \cite{constantin2004persistence}.
Its definition and methods have also found application to the description of economic data \cite{ren2003generalized}.
Some analytical and numerical predictions have been confirmed experimentally for liquid crystals \cite{yurke1997experimental}, diffusive systems \cite{wong2001measurement}, and fluctuating steps \cite{dougherty2002experimental}.
Most studies have focused on the persistence probability
\begin{equation}
  P(t)=1-\frac{N_f(t)}{V},
\end{equation}
where $V=L^d$ is the volume of the system and $N_f(t)$ quantifies the number of elements of the system that have had a first passage in the time interval $[0,t]$.
The first-passage time for the Ising model can be defined as the time at which the local order parameter changes its sign, i.e., a spin flips for the first time.
For many systems, one finds a power-law decay of the persistence probability as a function of time,
\begin{equation}
  \label{persistence}
P(t) \sim t^{-\theta}.
\end{equation}
For the persistence exponent $\theta$ there exist no exact estimates in $d=2$ dimensions, even for the NNIM \cite{bray2013persistence}.
For a quench from an uncorrelated starting configuration to $T=0$, numerically a value of $\theta=0.22$ \cite{derrida1994non,stauffer1994ising,manoj2000persistence} is found, whereas analytical approximations suggest $\theta \approx 0.19$ \cite{majumdar1996survival,sire2000analytical}.
In the more recent analysis of Ref.\ \cite{blanchard2014persistence} an influence of finite-time effects was recognized.
Incorporating these effects led to an estimate of $\theta=0.198(3)$, i.e., a value closer to the analytic approximation. 
Commonly, a value of $\theta=0.225$ is quoted \cite{ye2013nature,chakraborty2016fractality}.
\par
For some systems additional information may be extracted from investigating the correlation between persistent elements.
In Ref.\ \cite{manoj2000scaling} it was for example noted that a flip of a spin at a given site at time $t$ increases the chance of a neighboring spin flipping at time $t'>t$.
This naturally implies a spatial correlation of persistent spins.
They proposed to quantify these correlations by introducing the correlation function of persistent sites as
\begin{equation}
  D(r,t)=\frac{\langle \rho(x,t) \rho(x+r,t) \rangle}{\langle \rho(x,t) \rangle},
\end{equation}
where $\rho(x,t)=1$ if site $x$ is persistent and zero otherwise.
The angular brackets $\langle \ldots \rangle$ denote the average over initial conditions and independent trajectories.
With this definition, one has $\langle \rho(x,t) \rangle=P(t)$.
A length scale of this persistent lattice, $\ell_p(t)$, quantifying the separation of correlated and uncorrelated regions, grows akin to the length scale of the direct lattice $\ell(t)$ as a power-law function of time as
\begin{equation}
  \label{per_len}
  \ell_p(t) \sim t^{\alpha},
  \end{equation}
where $\alpha$ is again the growth exponent.
$D(r,t)$ has the dynamic scaling relation
\begin{equation}
  \label{DrtScalingPlot}
  D(r,t)=P(t)f(r/\ell_p(t)).
\end{equation}
For the scaling function $f(x)$ one finds
\begin{equation}
  \label{scalingDrt}
  f(x)=D(r,t)/P(t) \sim
  \begin{cases}
    x^{-\kappa} & x \ll 1\\
    1 & x \gg 1
  \end{cases},
\end{equation}
where $x=r/\ell_p(t)$ and $\kappa$ is an \emph{a priori} independent exponent.
For the NNIM it is known \cite{jain2000scaling} that the two-point correlator is independent of $t$ for $r \ll \ell_p(t)$.
A numerical confirmation that this is also true for the LRIM is presented below in Figs.~\ref{ScalingSig8}(a) and \ref{ScalingSig06Drt}(a) for $\sigma=8$ and $0.6$, respectively.
The only relation guaranteeing this is requiring $\ell_p^{-\kappa} \sim t^{-\theta}$.
This implies $D(r,t) \sim r^{-\kappa}$ for $\ell_p \gg r$ and plugging it into Eq.~\eqref{per_len} one arrives at the scaling relation 
\begin{equation}
\label{scaling}
\kappa \alpha=\theta.
\end{equation}
The exponent $\kappa$ is directly related to the fractal dimension of the persistent structures as can be seen from analytically analyzing the number of persistent spins in the square grid, which yields the relation \cite{manoj2000scaling,jain2000scaling,manoj2000persistence}
\begin{equation}
  \label{df_relation}
  d_f=d-\kappa.
\end{equation}
Combining Eqs.~\eqref{scaling} and \eqref{df_relation} one arrives at a relationship between the nonequilibrium exponents given as
\begin{equation}
  \label{relation_df_complete}
  d-d_f=\theta/\alpha.
\end{equation}
\par
Inserting the extreme estimates of $\theta$ into \eqref{scaling}, one arrives at an \textit{a priori} estimate of $1.55< d_f <1.62$.
These bounds are in agreement with all values estimated in the literature for $d_f$ \cite{jain2000scaling,chakraborty2016fractality}.
To conclude, for the NNIM one has: $\alpha=0.5$, $0.19<\theta<0.225$, $0.38 < \kappa < 0.45$, and $1.55< d_f <1.62$.
\par
For the LRIM with Hamiltonian \eqref{HLRIM} there exists a prediction of the asymptotic growth behavior reading $\ell(t) \sim t^{1/(1+\sigma)}$ for $\sigma<1$ and $\ell(t) \sim t^{1/2}$ for $\sigma>1$ with an additional multiplicative logarithmic correction at $\sigma=1$ \cite{Bray_only,bray1994growth,Rutenberg1994}.
This was derived using the deterministic time-dependent Ginzburg-Landau equation (without thermal contributions) and a continuous order parameter.
Numerically, this prediction has been confirmed in the $d=2$ LRIM \cite{christiansen2018,ConferenceChristiansen} for quench temperature $T = 0.1T_c \neq 0$ and subsequently in $d=1$ \cite{corberi2019one}.
In $d=1$ it was found in Ref.~\cite{corberi2019one} via a mapping to a two-domain approximation and onto the one-dimensional convection-diffusion equation that, apart from the above asymptotic growth law, one observes different initial growth regimes.
The scaling of the characteristic length in these three regimes depends on $\sigma$; at early times one observes a ``ballistic'' growth with $\ell(t) \sim t$, followed always by a regime where $\ell(t) \sim t^{1/(1+\sigma)}$, and finally only for $\sigma>1$ one has $\ell(t) \sim t^{1/2}$.
For long-range interacting systems in $d=1$ with any $\sigma$ at $T=0$ only spins at domain boundaries can flip and will, within the framework of a two domain approximation, always lead to a growth of the bigger domain since this is energetically favorable.
This means that at $T=0$ the ballistic regime becomes the asymptotic one and the other regimes are not observed.
We expect in $d=2$ at $T=0$ a similar and $\sigma$ \emph{independent} growth law, only the value of the growth exponent can not be derived as in $d=1$ and we do not expect to see exactly $\ell(t) \sim t$.
\par
For the investigation of the persistence in long-range interacting systems, only very little is known.
There exists one study \cite{ispolatov1999persistence} which investigates the case of zero-temperature coarsening in the $d=1$ LRIM modeled via Langevin dynamics which observes the above mentioned asymptotic prediction (and not the ballistic growth as in the NNIM) and $P(t) \sim \ell(t)^{-\overline{\theta}}$ with $\overline{\theta} \approx 0.17507588$ as for the nearest-neighbor time-dependent Ginzburg-Landau equation \cite{bray1994non}.
This corresponds trivially to $P(t) \sim t^{-\alpha\overline{\theta}}$, i.e., there a $\sigma$ dependent persistence exponent $\theta=\alpha\overline{\theta}$ was found.
For the one-dimensional NNIM, one instead finds $\theta=3/8$ \cite{derrida1994non,stauffer1994ising,derrida1995exponents,derrida1995exact,derrida1996exact}.

\begin{figure*}[!tbh]
  \newcommand{\mywidth}{0.28}
  \centering
  \begin{tabular}{ccc}
    $t=100$ & $t=200$ & $t=400$\\
    \includegraphics[width=\mywidth\textwidth]{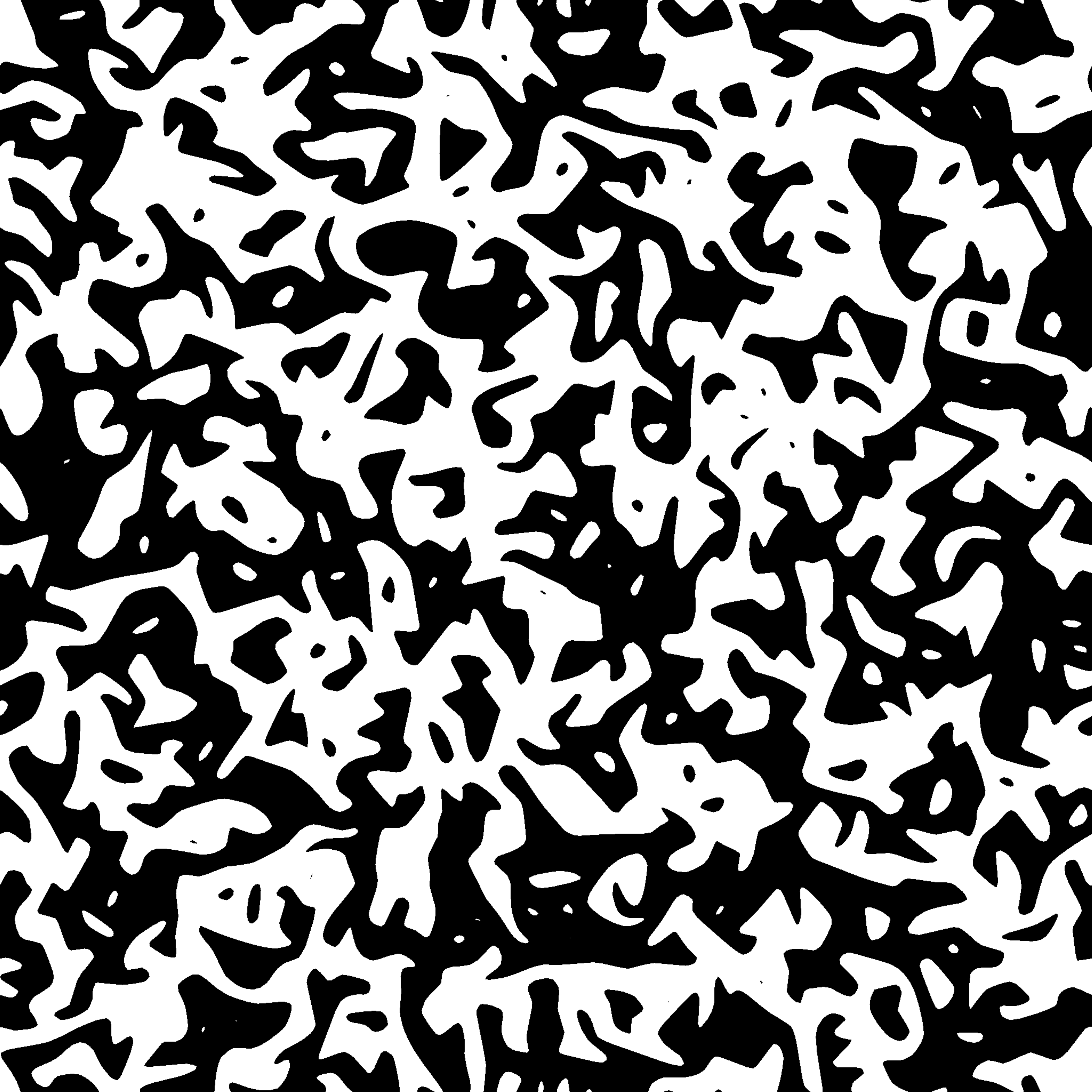} & \includegraphics[width=\mywidth\textwidth]{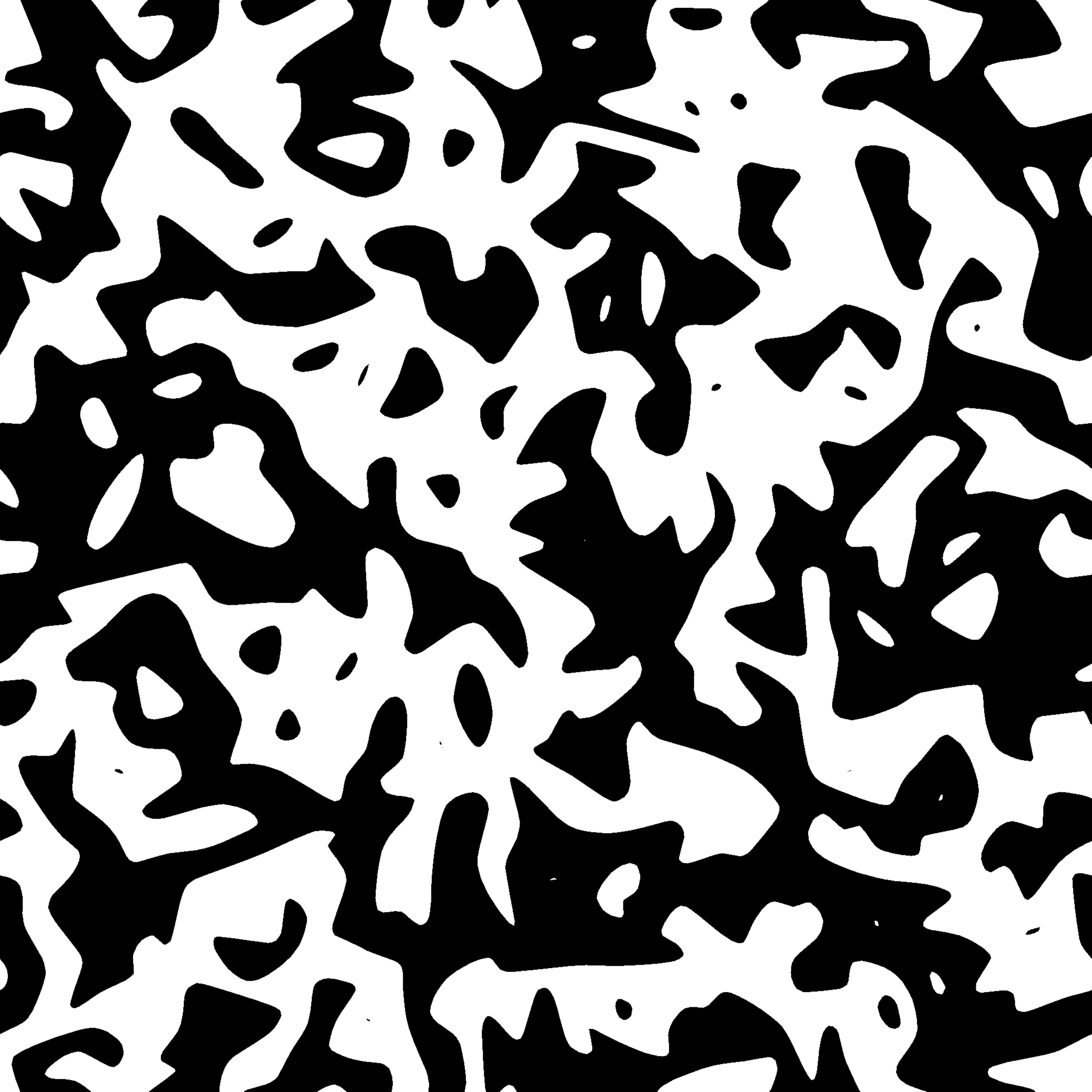} & \includegraphics[width=\mywidth\textwidth]{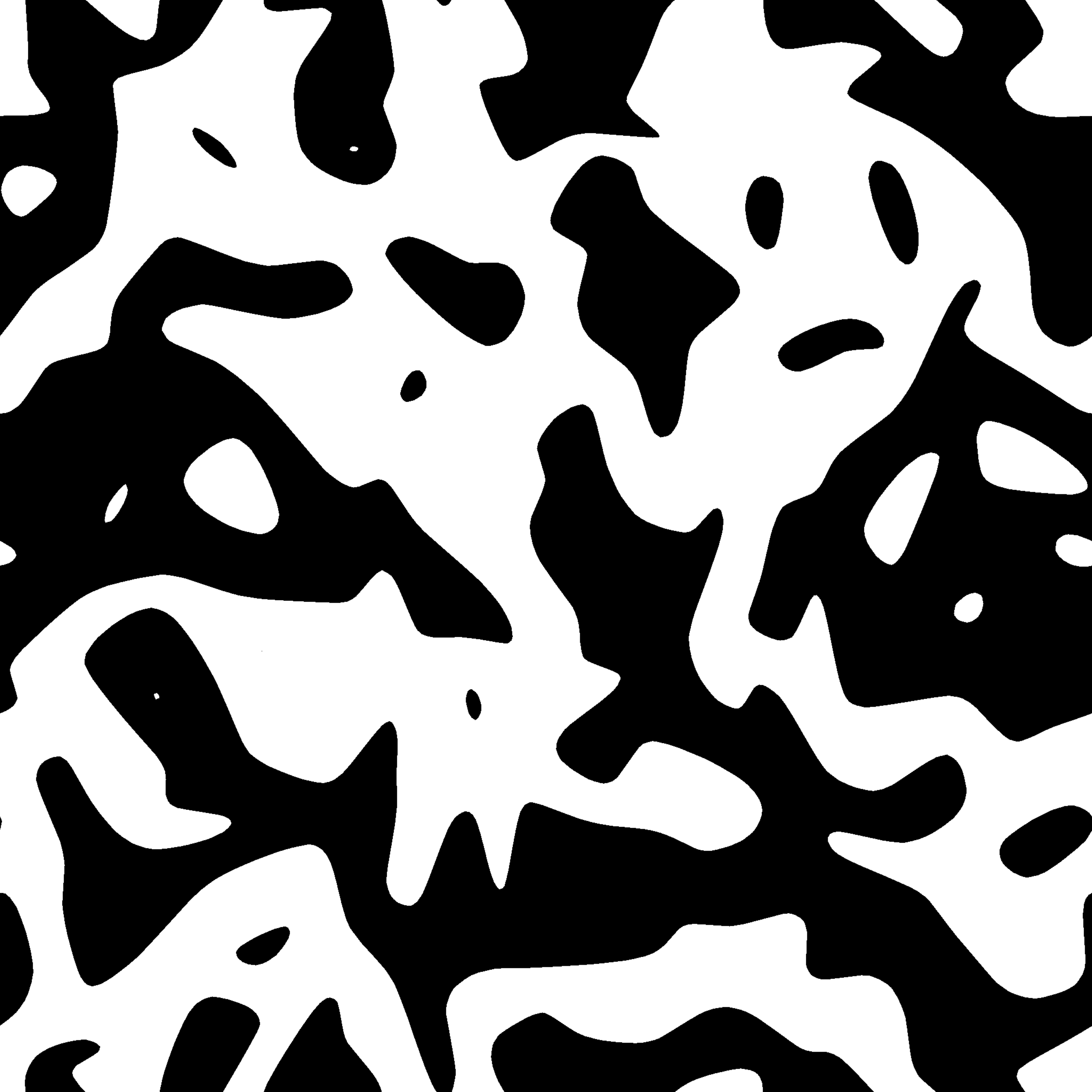} \\
    \includegraphics[width=\mywidth\textwidth]{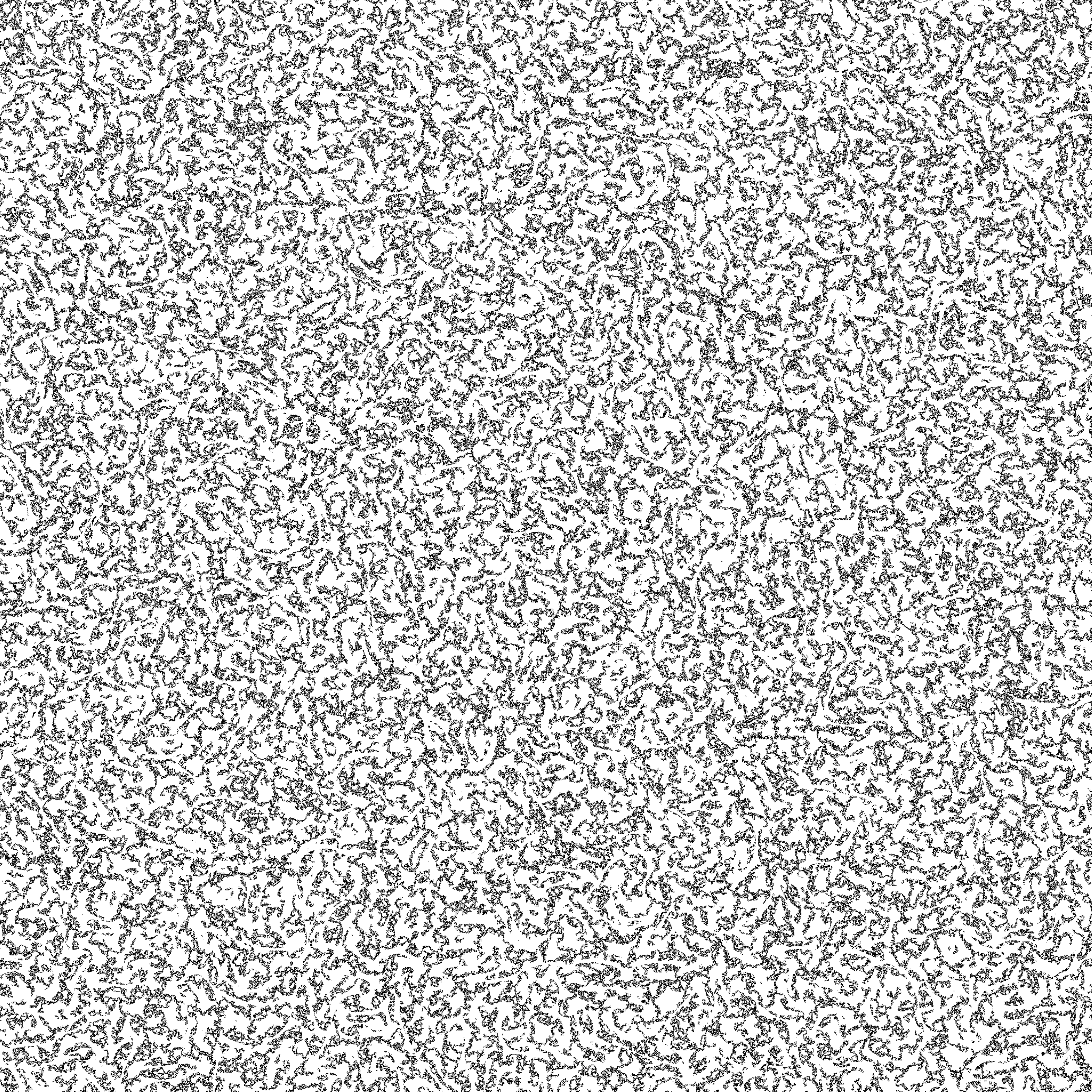} & \includegraphics[width=\mywidth\textwidth]{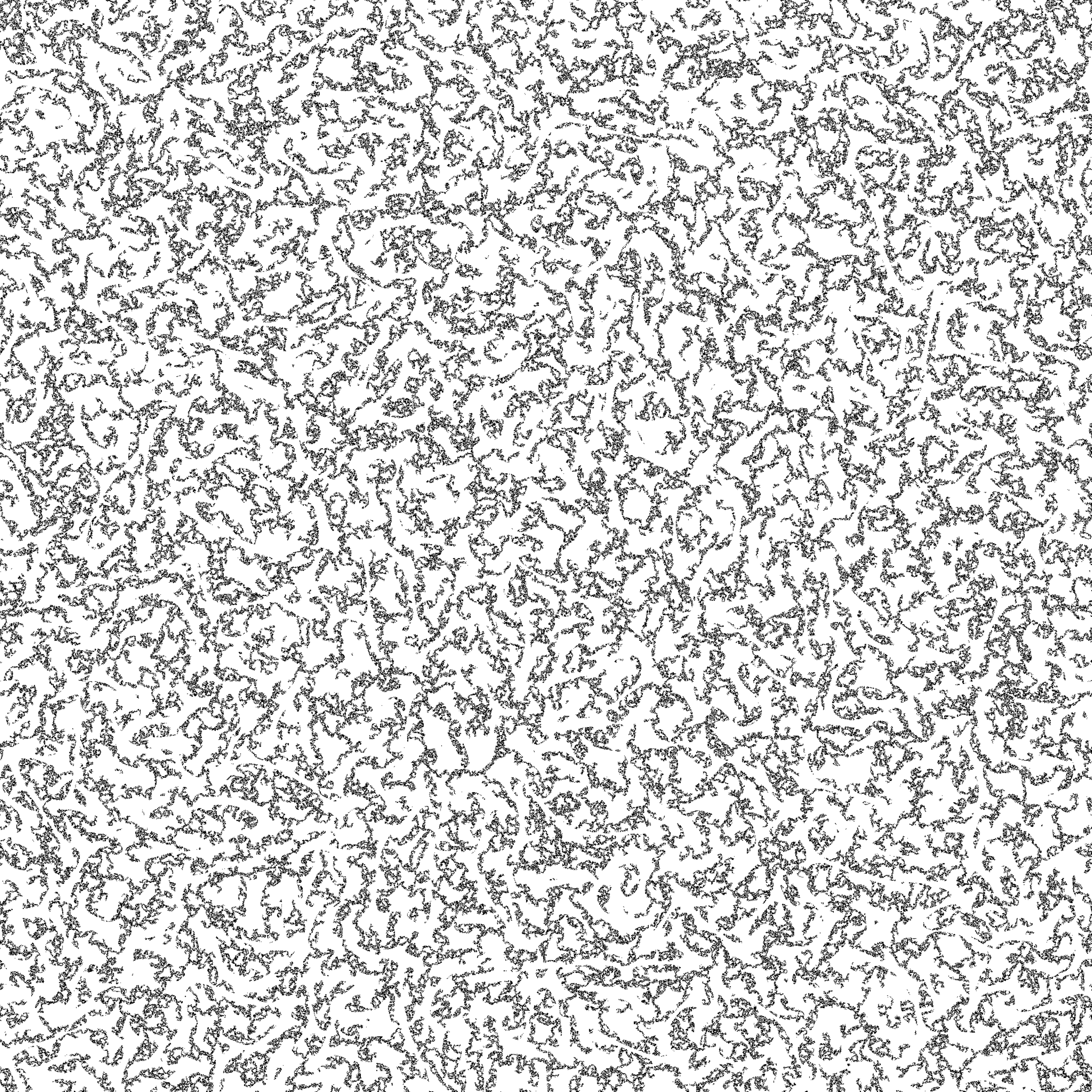} & \includegraphics[width=\mywidth\textwidth]{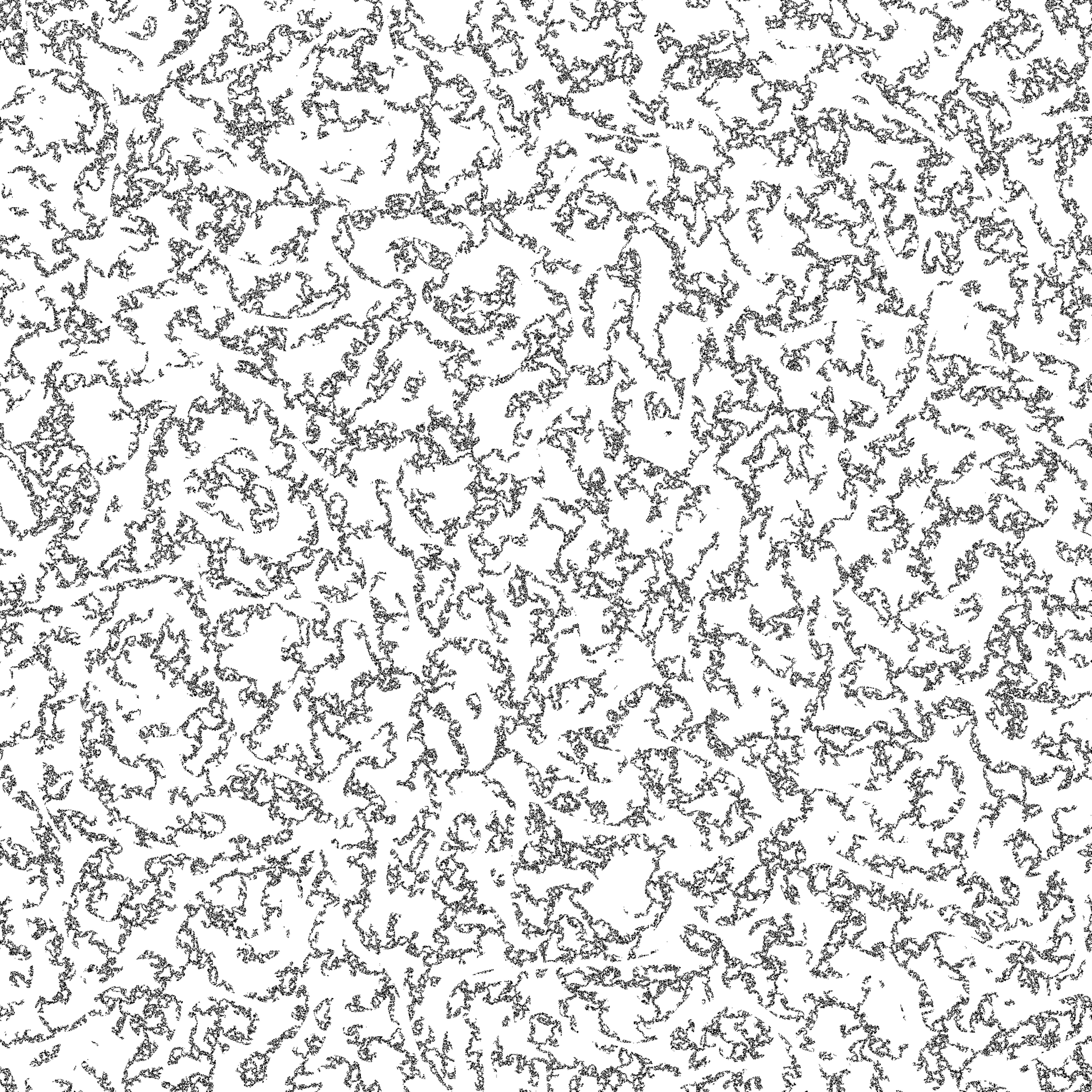} \\
  \end{tabular}
  \caption{Upper row: Configuration snapshots of the lattice after a quench from $T=\infty$ to $T=0$ for $\sigma=8$ and a system size of $L=2048$. The growth of ordered regions is apparent as the time $t=100$, $200$, $400$ increases. Lower row: Corresponding snapshots from the same simulation of the persistent lattice for the identical times. The fractal structure of these configurations is clearly visible.}
  \label{SnapSig8}
\end{figure*}

\section{Methods}
\label{Methods}
We simulate the LRIM with non-conserved order parameter in $d=2$ spatial dimensions with Hamiltonian \eqref{HLRIM} using a local Markov chain Monte Carlo algorithm.
When doing zero-temperature dynamics Monte Carlo, this algorithm degenerates to an energy minimization.
A randomly chosen spin is flipped if the flip results in a lower energy of the system.
If using the Glauber algorithm, a spin-flip resulting in zero energy change is accepted with $50$\% probability.
Using the Metropolis algorithm, such a spin-flip is always accepted.
For this work, we use the Glauber criterion, although in practice for systems with long-range interactions a proposed spin flip with zero energy change is very unlikely.
We define the unit of time as one sweep consisting of $L \times L$ spin flip attempts.
\par
When simulating systems with long-range interactions, the needed computational resources for a given system size are significantly increased when compared to the nearest-neighbor counterpart.
We recently proposed an approach where a simulation during a coarsening process is significantly sped up \cite{christiansen2018,ConferenceChristiansen}.
Additionally, finite-size effects are very prominent in long-range interacting systems.
Therefore we use Ewald summation to calculate the effective interactions $J_{i,j}$ between spin $i$ and $j$ \cite{ewald1921berechnung,horita2017upper}, which are said to reduce the effects of a finite system.
Nonetheless, this still makes a careful treatment of finite-size effects in systems with strongly long-range interactions necessary, which we will put an emphasis on in Section \ref{LR}.
\par
The length $\ell(t)$ entering the asymptotic scaling law \eqref{len} is extracted by taking the intersection of the two-point equal-time correlation function
\begin{equation}
C(r,t)=\langle s_i(t)s_j(t) \rangle - \langle s_i(t) \rangle \langle s_j(t) \rangle 
\end{equation}
with a reasonable choice of a constant value in the range $(0,1)$ (here we chose $0.5$).
The value of $r$ at this intersection is then interpreted as the characteristic length $\ell(t)$ at this time.
When plotting self-consistently $C(r,t)$ versus $r/\ell(t)$, due to this being a scaling process, one expects the curves for all times $t$ to collapse in the scaling regime.
The calculation of this correlation function is numerically sped up via a fast Fourier transform.
The length of the persistent structures $\ell_p(t)$ is, somewhat similarly, extracted by determining $r$ where $D(r,t)$ first crosses $P(t)$.
\par
All simulation results, apart from the snapshots, are obtained by averaging over at least $40$ independent runs.
This is realized by running the simulation with otherwise identical parameters for different random number generator seeds, corresponding to an average over different initial conditions and time evolutions.
For reference, a single run of system size $L=4096$ takes roughly $3.5$ weeks on $20$ cores when parallelized with OpenMP.
\par
For our fits, we make use of Jackknifing, i.e., when we have $N$ independent runs, we also perform $N$ fits each on data sets containing the information from $N-1$ runs.
This allows us to calculate (reliable) error bars on the fit parameters by taking care of the (trivially) introduced correlation by using the same data many times \cite{efron1982jackknife}.

\section{Results}
Naively, one would expect to recover all exponents as in the NNIM for $\sigma \rightarrow \infty$.
As we have already discussed in the introduction, however, we strongly suspect that the growth exponent $\alpha$ will indeed not be recovered.
To check the specifics and for estimating the value of $\alpha$, we first focus on the case of $\sigma=8$, which for quenches to $T \neq 0$ \cite{christiansen2018,christiansen2019non} and in equilibrium \cite{fisher1972critical,sak1973recursion} undoubtedly corresponds to the short-range-like regime of the long-range interacting model.
\subsection{Large $\sigma$}
\label{SR}
In Fig.~\ref{SnapSig8} we first show snapshots of the direct (upper row) and persistent (lower row) lattice for $t=100$, $200$, $400$ from a single quench for $\sigma=8$ and system size $L=2048$.
As expected, we observe the growth of ordered structures.
Here, we want to note a difference between the NNIM and this model.
While for the NNIM roughly $1/3$ of the simulations at zero temperature get stuck in (in principle meta stable) stripe-like configurations \cite{safran1983kinetics,spirin2001fate,spirin2001freezing,blanchard2013frozen}, this is much less likely in the LRIM.
\begin{figure}
  \includegraphics{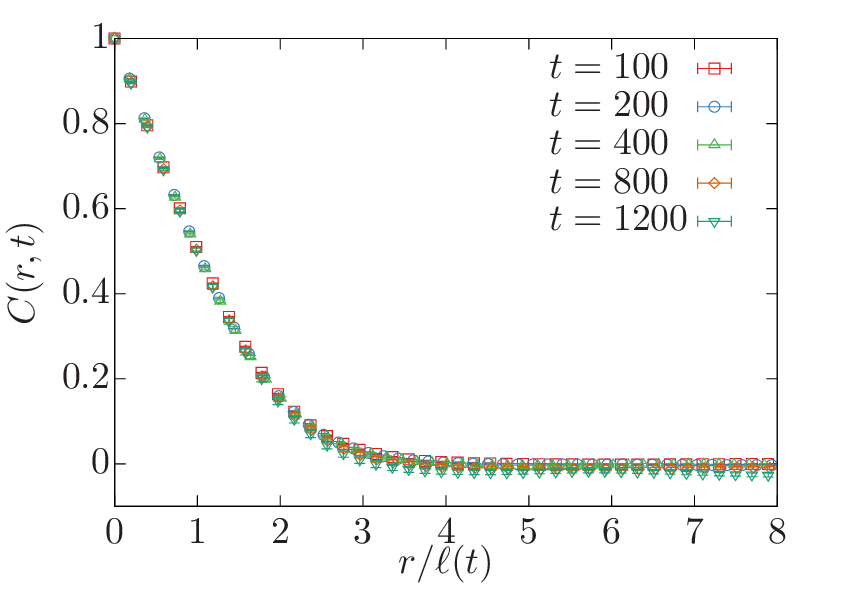}
  \caption{Correlation function of the direct lattice $C(r,t)$ for $\sigma=8$ and $L=2048$ plotted against distance $r$ scaled by the characteristic length $\ell(t)$ extracted from the intersection of this correlation function with $0.5$.}
  \label{directCorFig}
\end{figure}

\begin{figure}
  \includegraphics{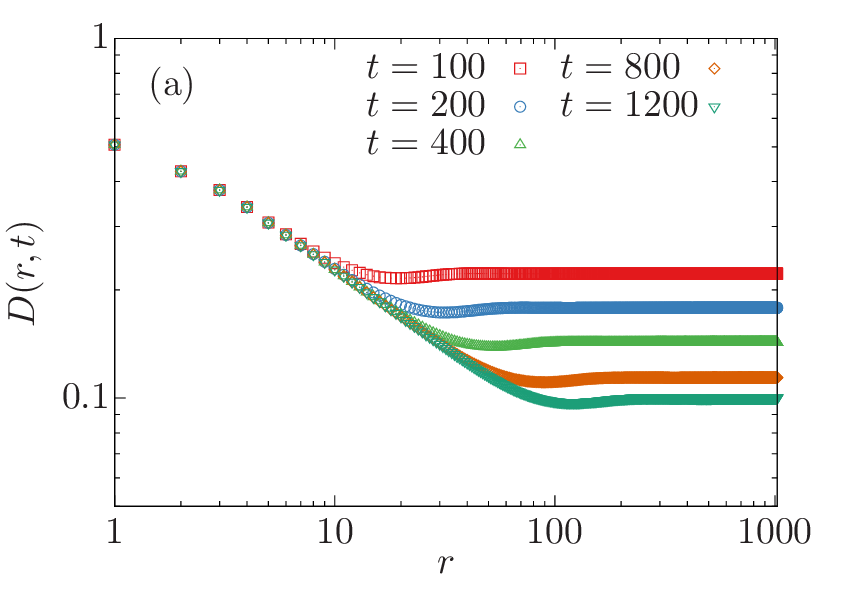}
  \includegraphics{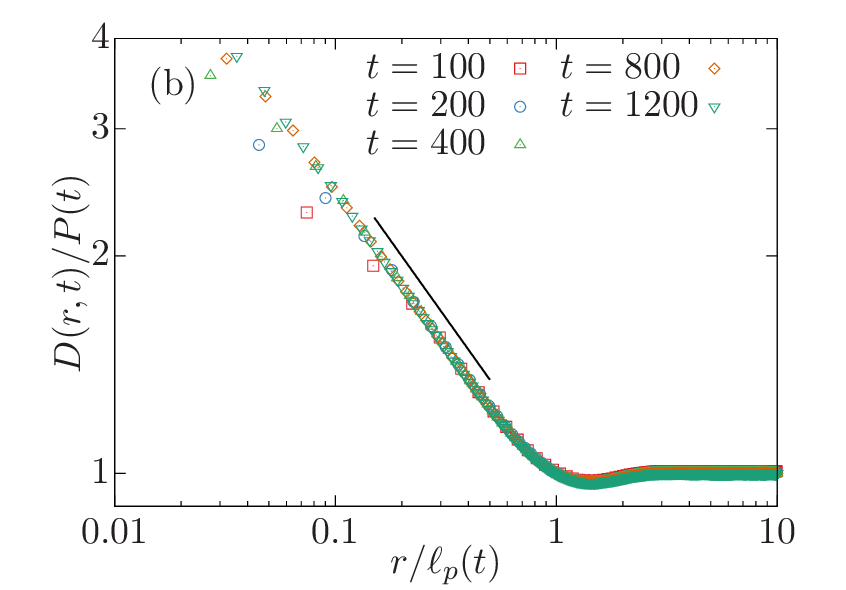}
  \caption{(a) Correlation function of the persistent lattice $D(r,t)$ against distance $r$ for $\sigma=8$ and times $t=100,200,400,800,1200$. (b) Same data as in (a), but $D(r,t)$ scaled by $P(t)$ now plotted against the scaled distance $r/\ell_p(t)$, as described in Eq.~\eqref{scalingDrt}. The solid line in (b) corresponds to a power law with exponent $\kappa=d-d_f=0.430$.}
  \label{ScalingSig8}
  \end{figure}
To be able to investigate any growth laws, we have to establish that the data for the correlation functions $C(r,t)$ and $D(r,t)$ fall on a master curve when properly rescaled.
This exercise is shown in Fig.~\ref{directCorFig} for the correlation function of the direct lattice $C(r,t)$.
In Fig.~\ref{ScalingSig8}(a) we first present the unscaled correlation function of the persistent lattice $D(r,t)$ for times $t=100,200,400,800,1200$, confirming that $D(r,t)$ does not depend on $t$ for $r \ll l_p(t)$.
In Fig.~\ref{ScalingSig8}(b) we show for the same data the corresponding scaled plot according to Eq.~\eqref{DrtScalingPlot}.
Both correlation functions collapse well; dynamic scaling is found and both $\ell(t)$ and $\ell_p(t)$ are properly estimated.
We perform fits of the power-law part of Eq.~\eqref{scalingDrt}, where error bars are estimated using Jackknifing.
The data considered is for $t=1200$ (the largest time plotted in Fig.~\ref{ScalingSig8}(b)) in the range $0.15 < r/\ell_p(t) < 0.5$.
We find $\kappa=0.430(2)$ and $\chi^2_r=1.9(9)$, which is shown as a solid line in Fig.~\ref{ScalingSig8}(b).
Here, the reduced chi-square $\chi^2_r=\chi^2/\mathrm{DOF}$ measures the goodness of fit, where $\mathrm{DOF}$ denotes the degrees of freedom.
This value of $\kappa$ corresponds via Eq.~\eqref{df_relation} to $d_f=1.570(2)$ and is in good agreement with the results observed in the NNIM \cite{jain2000scaling,chakraborty2016fractality}.
We want to stress that this estimate is not to be taken too seriously, because the data is well compatible with a relatively wide range of $\kappa$ when only slightly adjusting the fitting range ($0.42 < \kappa < 0.44$ cannot be unambiguously ruled out), corresponding to $1.56<d_f<1.58$.
\begin{figure}
\includegraphics{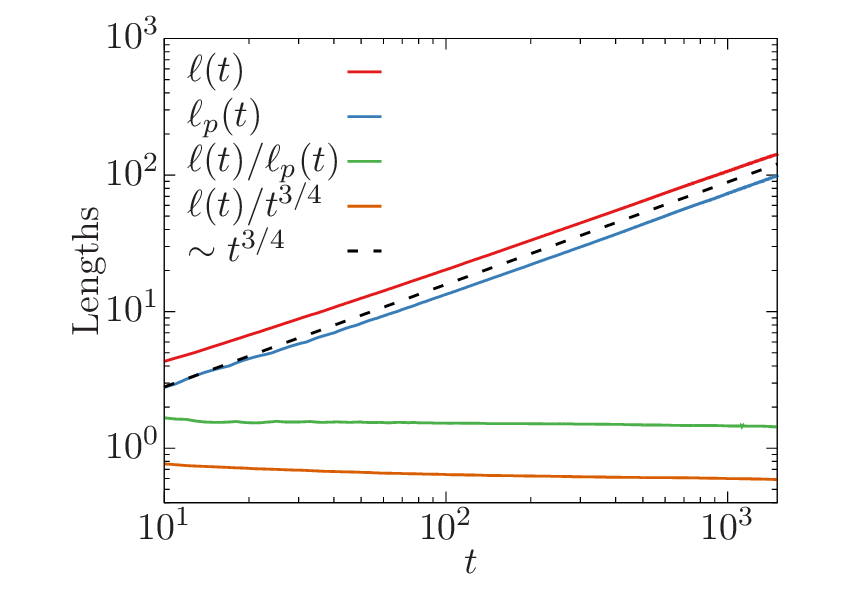}
\caption{
  The relevant lengths $\ell(t)$ and $\ell_p(t)$ for $\sigma=8$ and $L=2048$.
  The statistical errors are of the order of the line width.
  The dashed black line corresponds to a power law in time $t$ with growth exponent $\alpha=3/4$.
  Additionally plotted is $\ell(t)/\ell_p(t)$ which is expected to be constant as both lengths follow the same power law and $\ell(t)/t^{3/4}$ which is also roughly constant in a relatively long time window, indicating that the growth follows $\ell(t)\sim t^{3/4}$.}
\label{length_sig_8}
\end{figure}
\begin{figure}
  \includegraphics{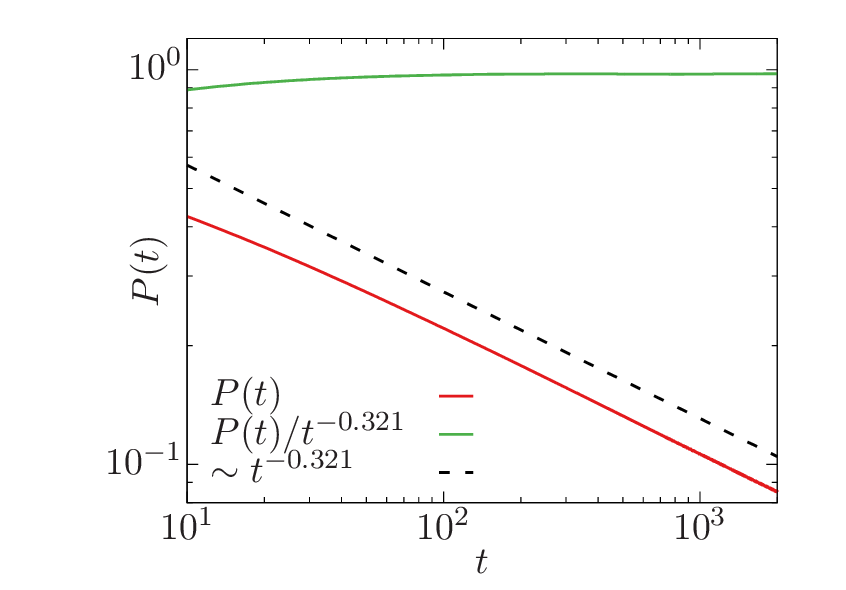}
  \caption{Persistence probability $P(t)$ versus time $t$ for $\sigma=8$ and $L=2048$. The dashed black line corresponds to a power-law decay with exponent $\theta=0.321$.
    To confirm that this exponent is compatible with the data, we have additionally plotted $P(t)/t^{-\theta}$.}
  \label{PersistenceSig8}
\end{figure}

\begin{figure*}
    \newcommand{\mywidth}{0.28}
  \centering
  \begin{tabular}{ccc}
    $t=100$ & $t=200$ & $t=400$\\
    \includegraphics[width=\mywidth\textwidth]{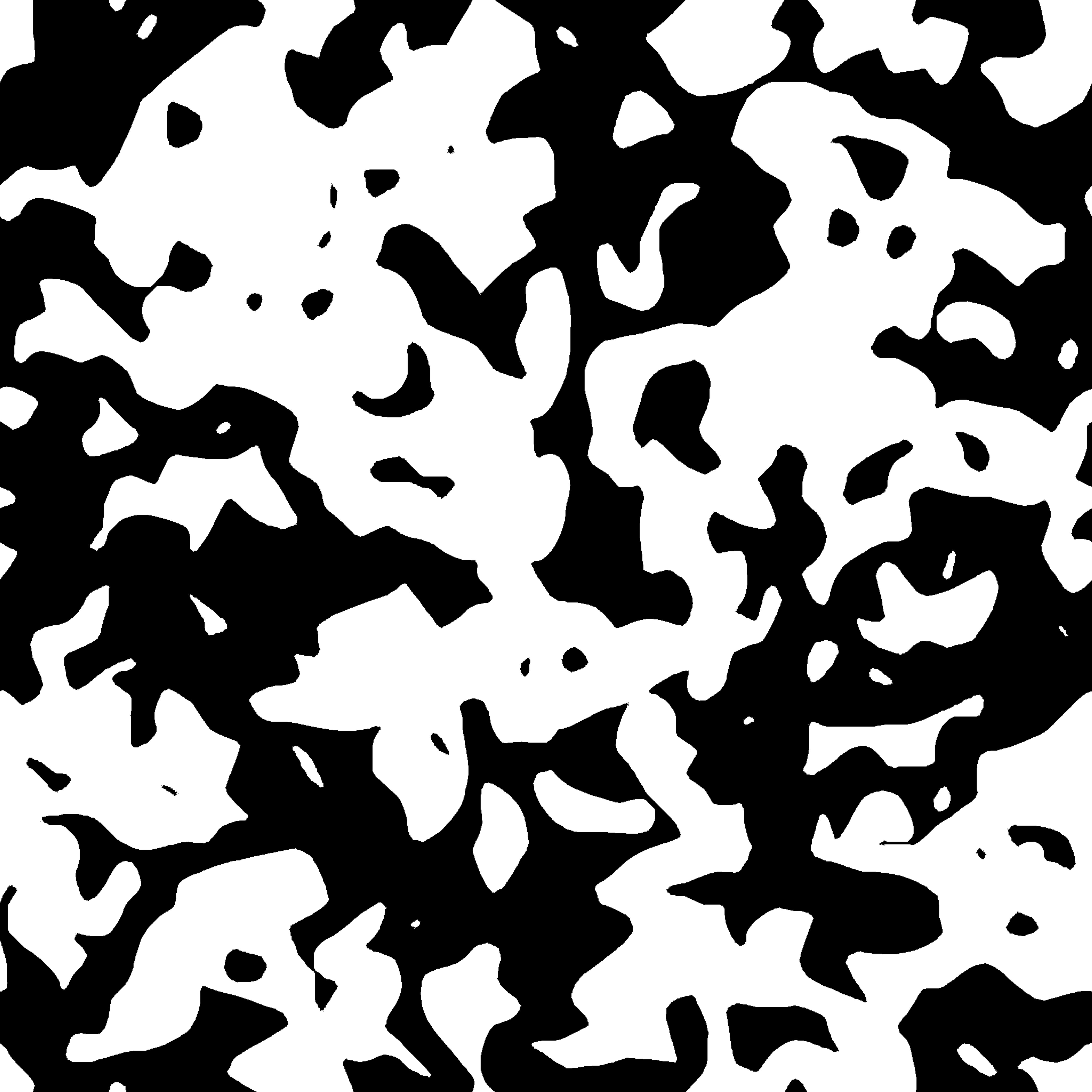} & \includegraphics[width=\mywidth\textwidth]{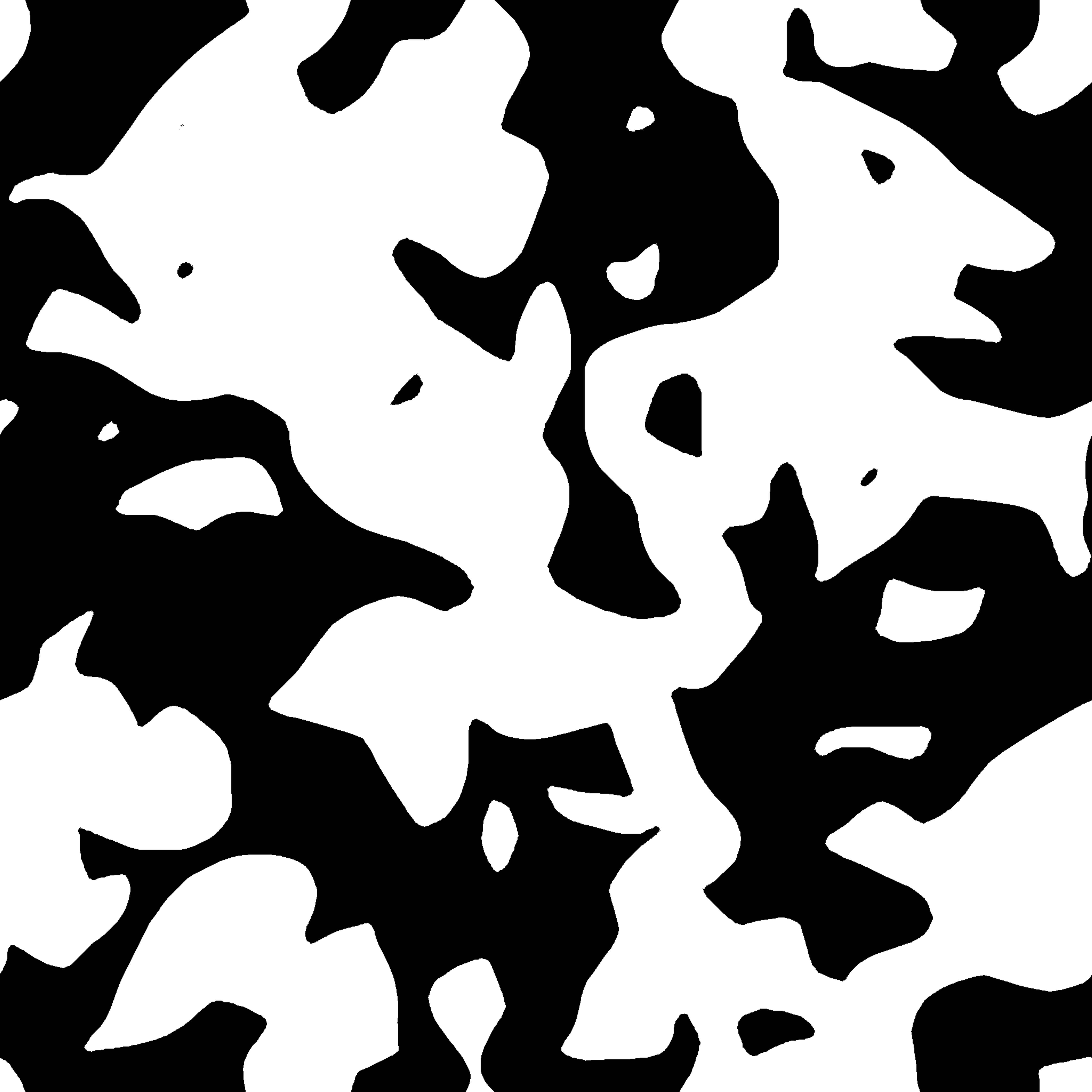} & \includegraphics[width=\mywidth\textwidth]{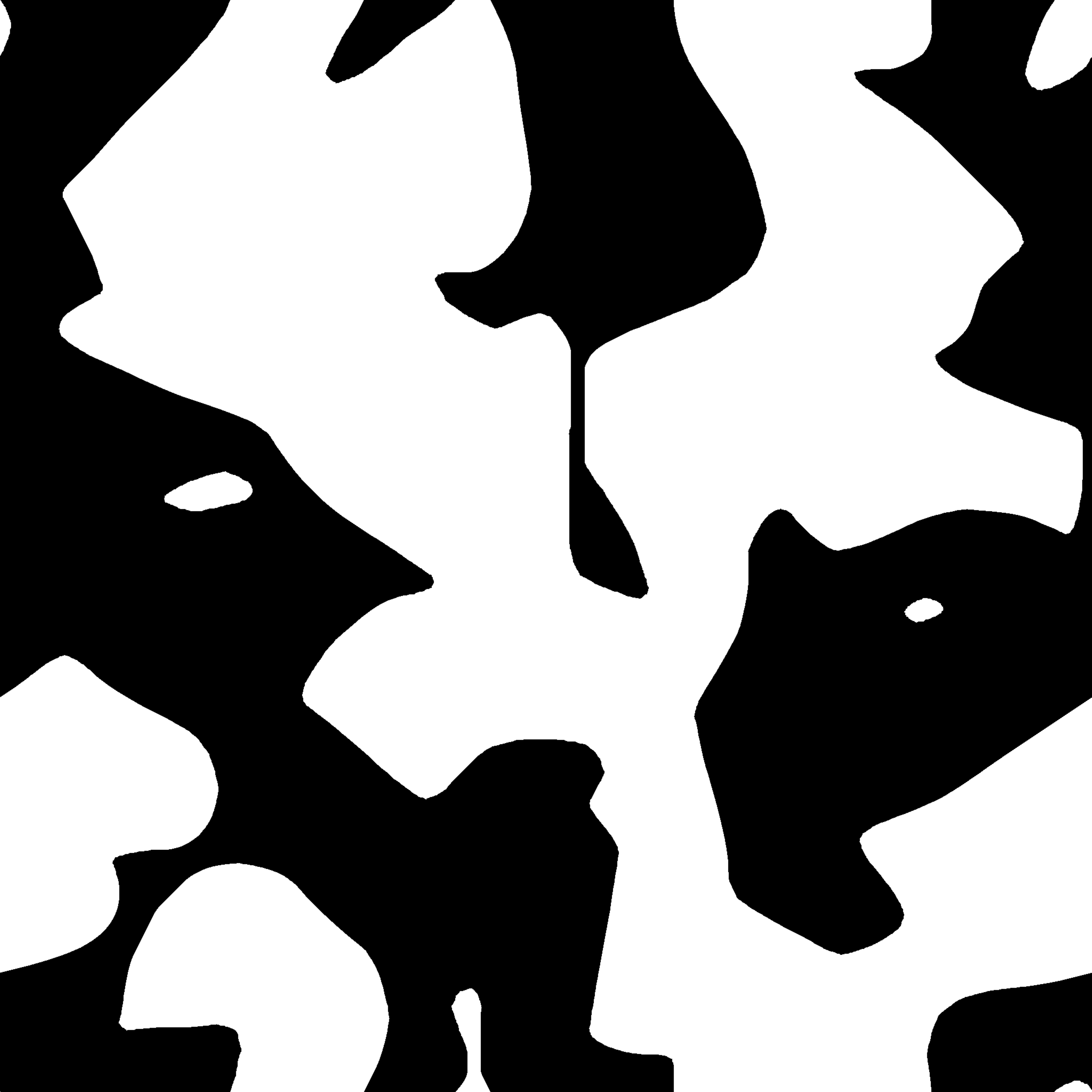} \\
    \includegraphics[width=\mywidth\textwidth]{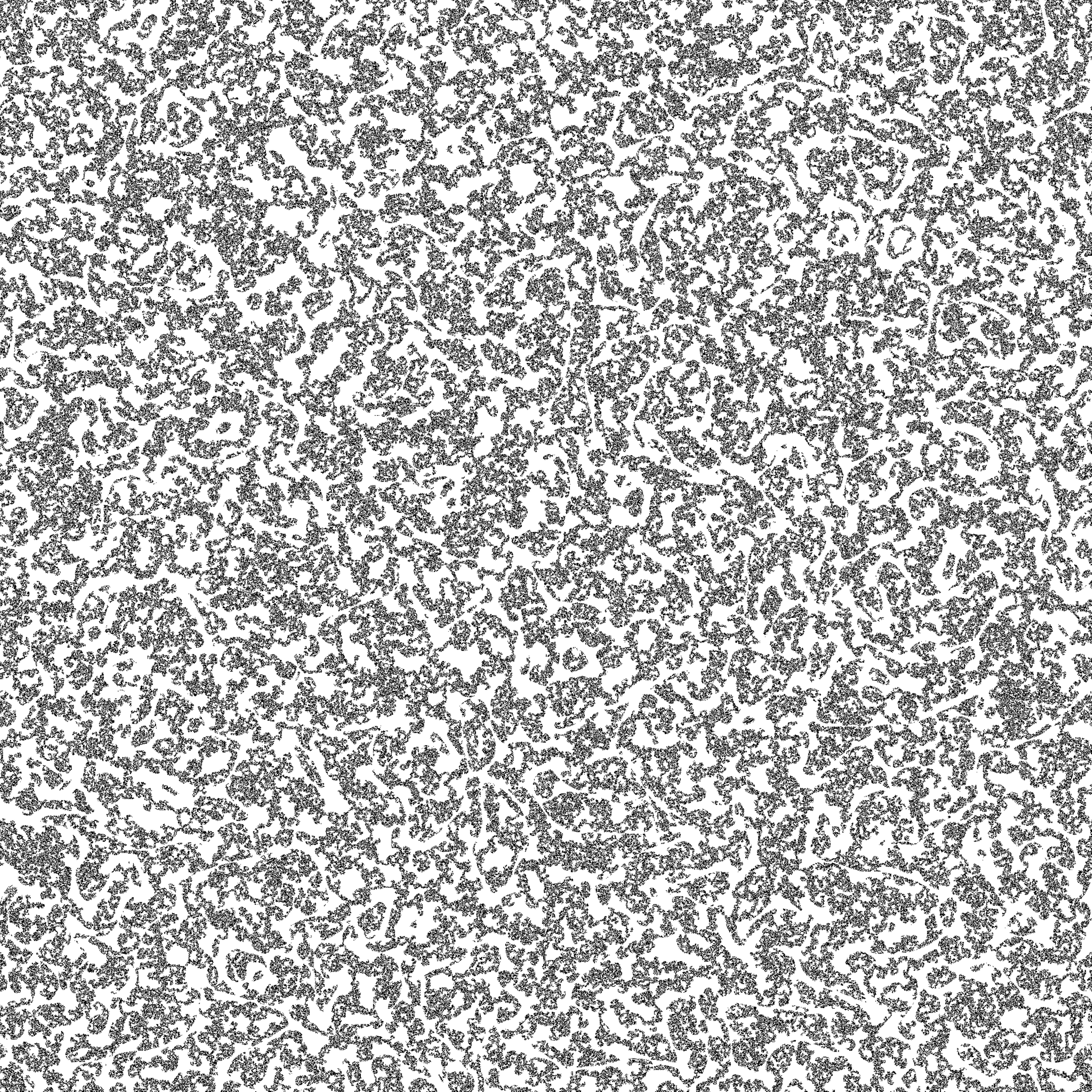} & \includegraphics[width=\mywidth\textwidth]{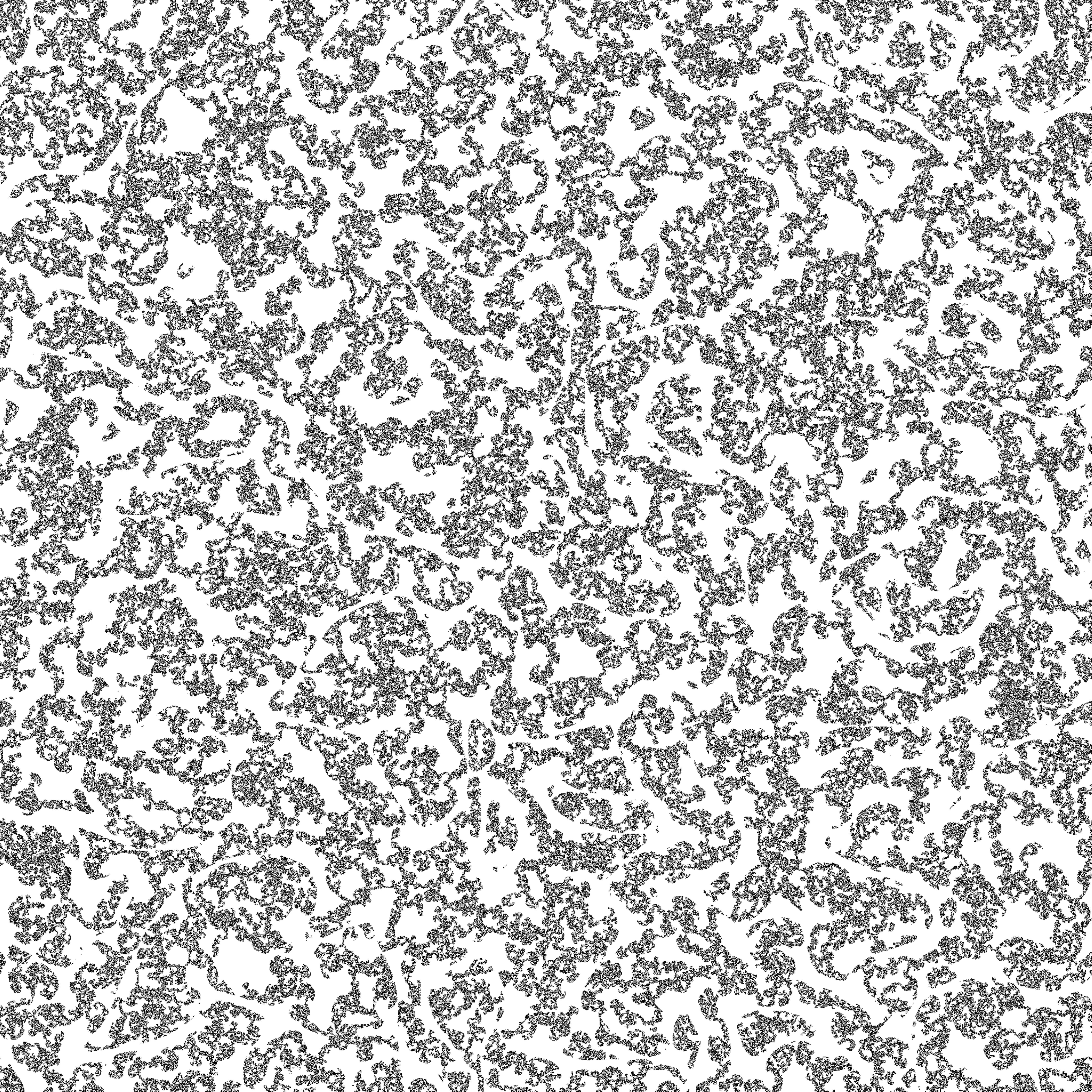} & \includegraphics[width=\mywidth\textwidth]{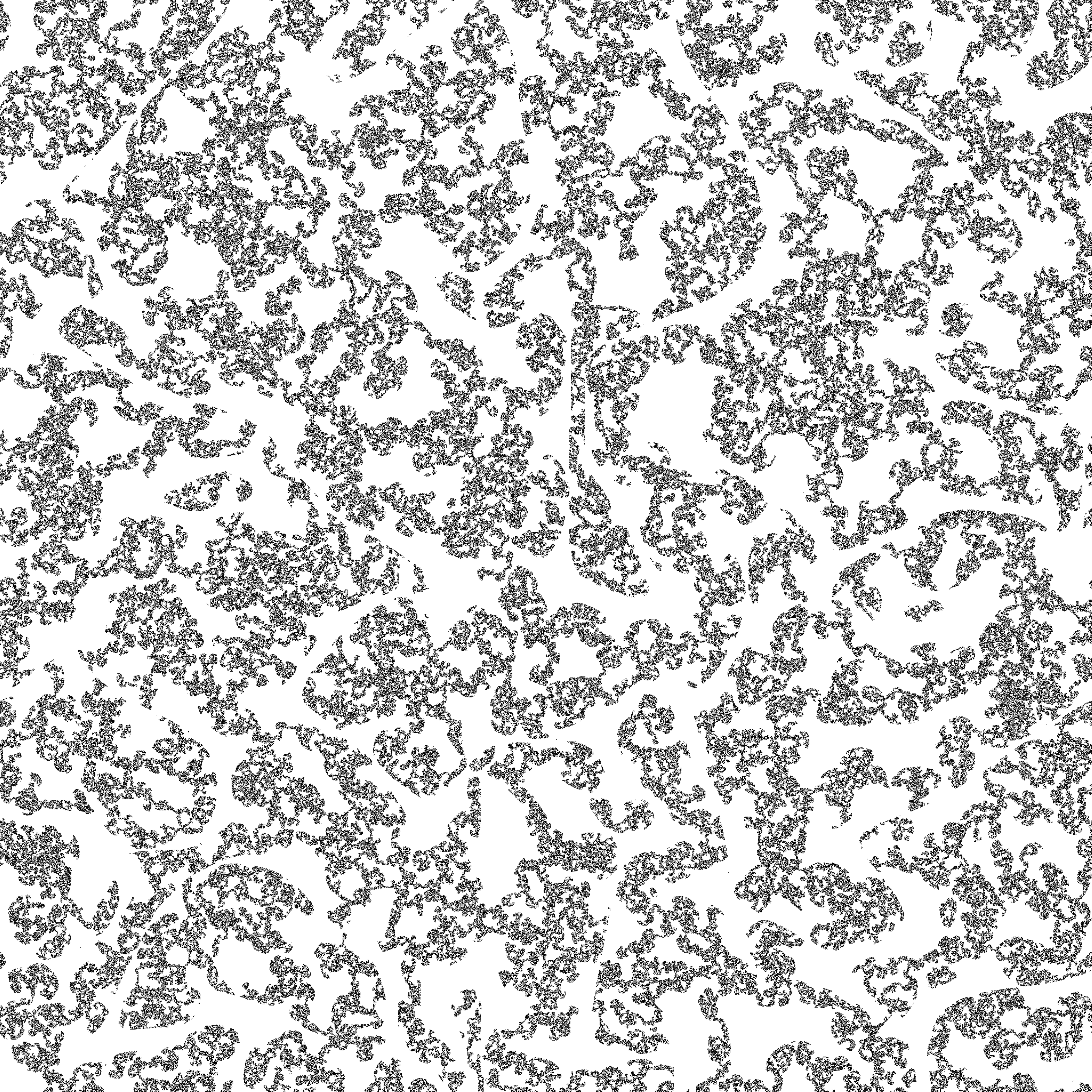} \\
  \end{tabular}
  \caption{Upper row: Configuration snapshots of the lattice after a quench from $T=\infty$ to $T=0$ for $\sigma=0.6$ and a system size of $L=2048$.
    The growth of ordered regions is apparent as the time $t=100$, $200$, $400$ increases. Lower row: Corresponding snapshots from the same simulation of the persistent lattice for the identical times. The fractal structure of these configurations is clearly visible.}
  \label{SnapSig06}
\end{figure*}
\begin{figure}
  \includegraphics{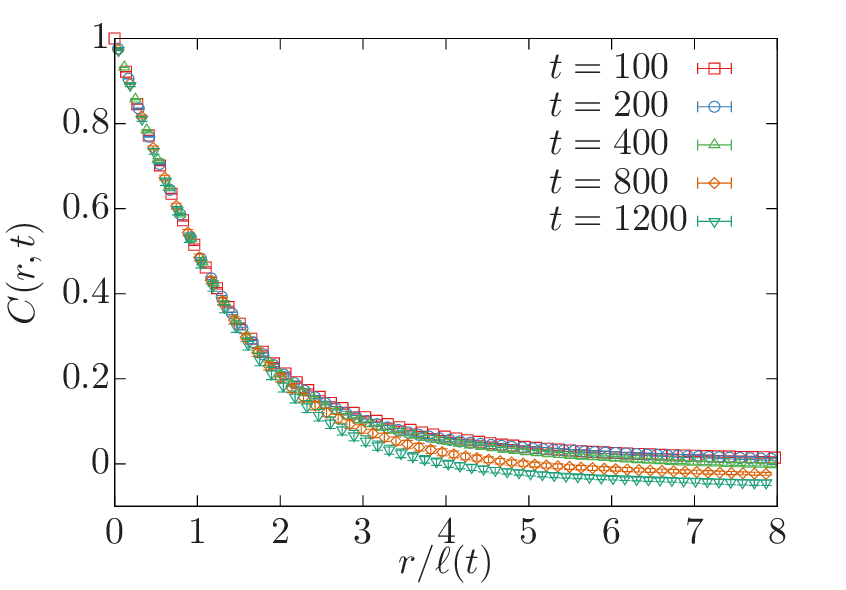}
  \caption{Correlation function of the direct lattice $C(r,t)$ for $\sigma=0.6$ and $L=4096$ plotted against distance $r$ scaled by the characteristic length $\ell(t)$ extracted from the intersection of this correlation function with $0.5$ for times $t=100,200,400,800,1200$.}
  \label{ScalingSig06}
\end{figure}

\begin{figure}
  \includegraphics{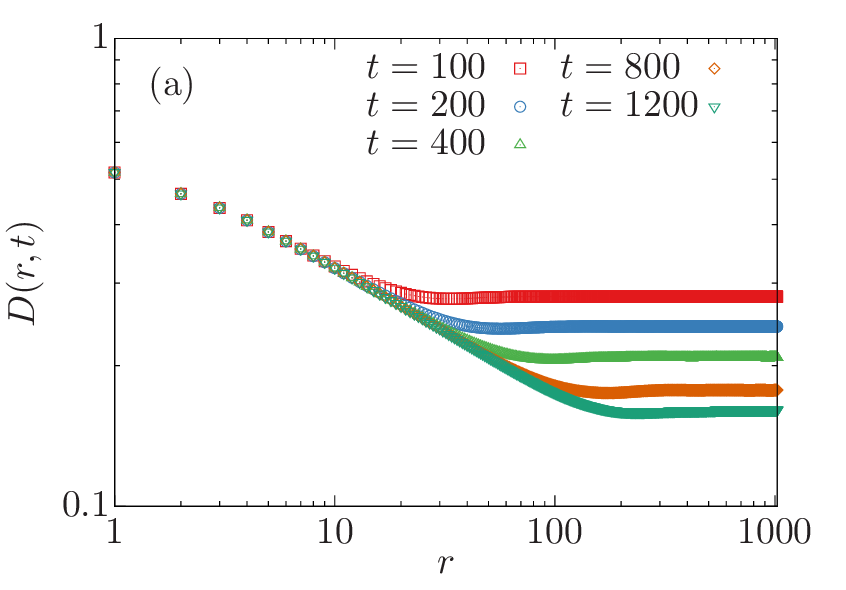}
  \includegraphics{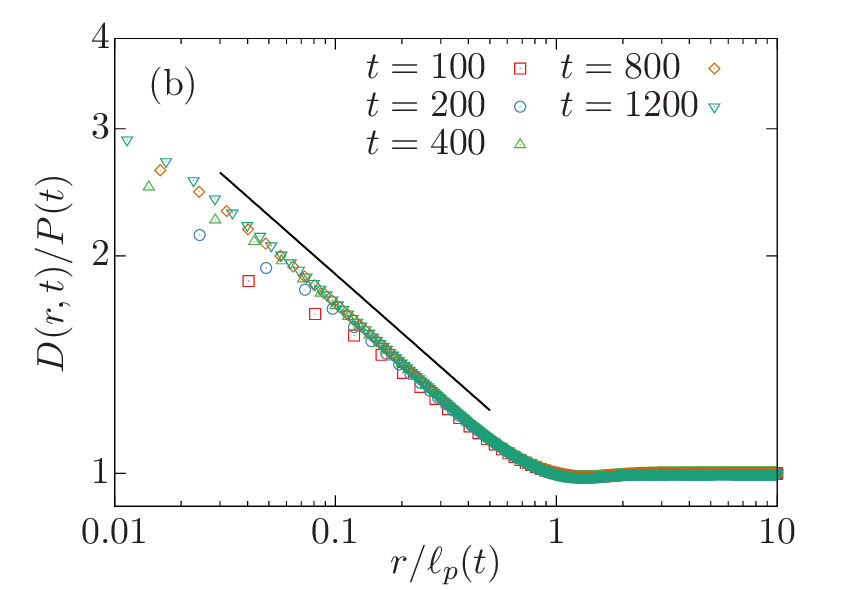}
  \caption{(a) Correlation function of the persistent lattice $D(r,t)$ against distance $r$ for $\sigma=0.6$, $L=4096$, and times $t=100,200,400,800,1200$. (b) Same data as in (a), but $D(r,t)$ scaled by $P(t)$ now plotted against the scaled distance $r/\ell_p(t)$, as described in Eq.~\eqref{scalingDrt}. The solid line in (b) corresponds to a power law with exponent $\kappa=d-d_f=0.268$.}
  \label{ScalingSig06Drt}
\end{figure}
\begin{figure}
  \includegraphics{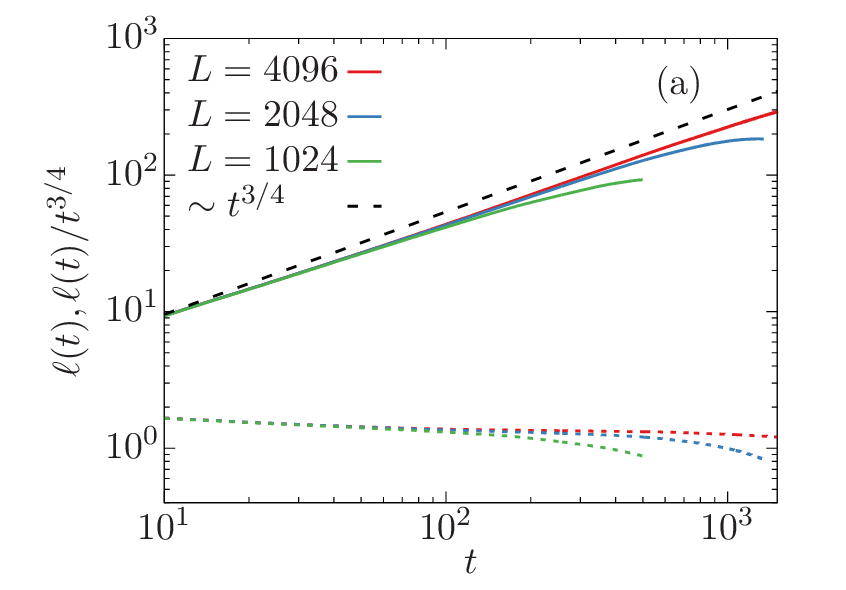}
  \includegraphics{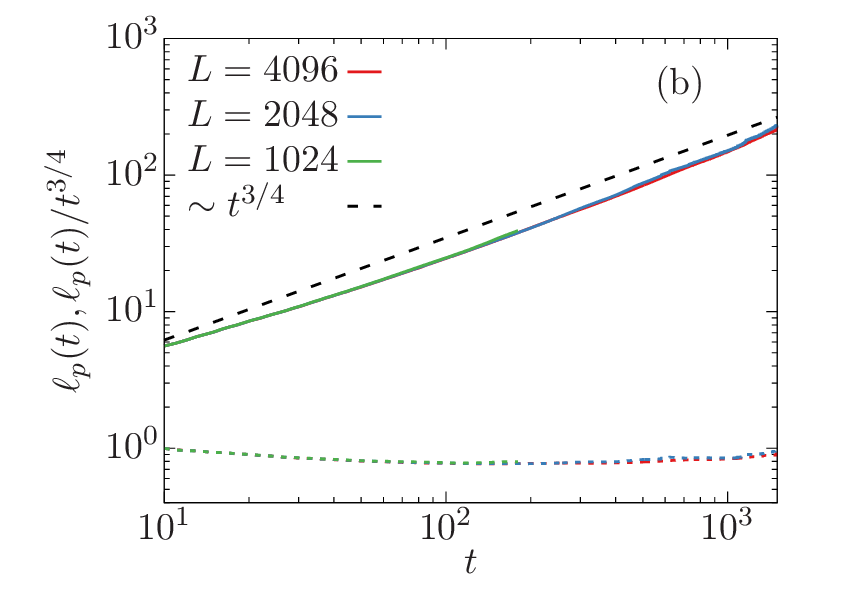}
  \caption{(a) Characteristic length $\ell(t)$ for $\sigma=0.6$ and $L=1024$, $2048$, and $4096$ as solid colored lines.
    The statistical errors are of the order of the line width.
    The dashed black line corresponds to a power law in time $t$ with growth exponent $\alpha=3/4$.
    Additionally plotted is $\ell(t)/t^{3/4}$ for all system sizes as dashed colored lines, which for the correct scaling law should approach a constant.
    (b) Similar as (a), but for $\ell_p(t)$.}
\label{length_sig_06}
\end{figure}
\begin{figure}
  \includegraphics{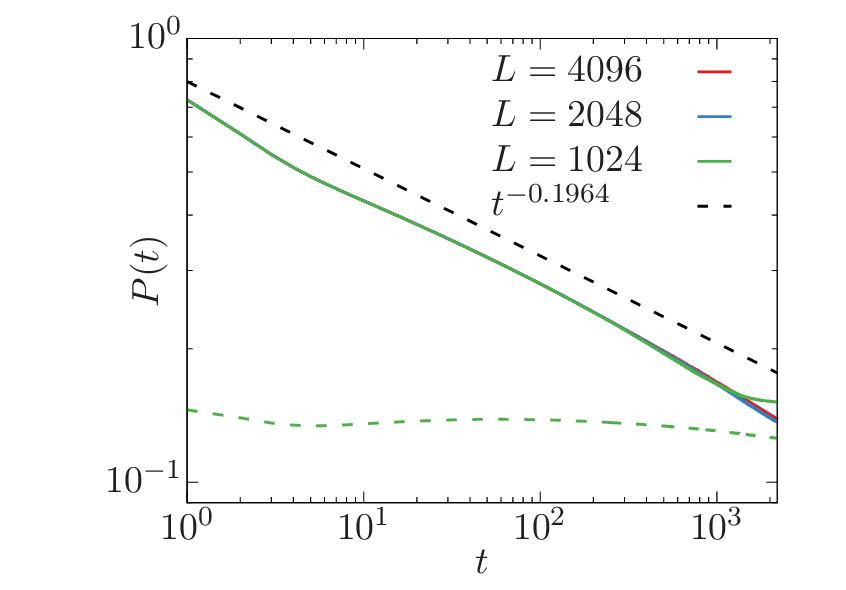}
  \caption{
    Persistence probability $P(t)$ versus time $t$ for $\sigma=0.6$ and $L=1024,2048,4096$.
    The dashed black line corresponds to a power-law decay with exponent $\theta=0.1964$.
    To confirm that this exponent is compatible with the data, we have additionally plotted $P(t)/t^{-\theta}$ as dashed green line for $L=4096$.}
  \label{PersistenceSig06}
\end{figure}
\par
We next investigate in Fig.~\ref{length_sig_8} both the length scales $\ell(t)$ and $\ell_p(t)$ already used as a rescaling factor.
We expect both of them to follow the same power law with growth exponent $\alpha$.
In Fig.~\ref{length_sig_8}, both lengths are plotted against time $t$ for $\sigma=8$ and $L=2048$.
To extract a numerical estimate of $\alpha$, we perform fits in the range $10^2 \le t \le 10^3$, giving $\alpha=0.724(6)$ with $\chi^2_r \approx 0.4$ for $\ell$ and $\alpha = 0.743(9)$ with $\chi^2_r \approx 1$ for $\ell_p$.
These values are suggestively close to $\alpha=3/4$, so that we plot as a dashed line a power law with exponent $\alpha=3/4$, which appears to be compatible with the data for both lengths.
To further consolidate this claim numerically, we plot $\ell(t)/t^{\alpha}$, which should be constant in the scaling region if this is the correct exponent.
In the interval $[10^2,10^3]$, this very much is the case, as could also be appreciated from a plot on a linear scale (not shown).
This confirms our expectation that one should observe at zero temperature in the LRIM a different growth exponent $\alpha$ than in the NNIM and the LRIM quenched to $T \neq 0$.
It remains to be checked, whether also at $T=0$ this exponent is $\sigma$ independent, which will be investigated in the next sections.
Apart from this empirical evidence, however, we have at the time being no simple theoretical explanation for this rational exponent.
Finally, to confirm that both lengths follow the same power law, we plot $\ell(t)/\ell_p(t)$ which correspondingly should be constant as well in this region.
Within the available data accuracy this can be again easily appreciated from Fig.~\ref{length_sig_8}.
\par
Finally, the persistence probability $P(t)$ is presented in Fig.~\ref{PersistenceSig8}.
To check the validity of Eq.~\eqref{scaling} we calculate the value of $\theta \approx 0.319$ from the fitted values, using $\alpha$ extracted from $\ell_p(t)$.
We also perform a fit in the range of $10^2 \le t \le 10^3$ as before and find $\theta = 0.321(2)$.
By varying the fit range, we cannot rule out a slightly different exponent $0.31<\theta<0.34$, however, the numeric value is very suggestive of the ``nice'' rational exponent $1/3$ for $\sigma \rightarrow \infty$.
Inserting $\theta=1/3$ and $\alpha=3/4$ into Eq.~\eqref{scaling} gives a value of $\kappa=4/9=0.44\overline{4}$ or $d_f=14/9=1.55\overline{5}$.
This value appears to be approximately compatible with the estimated value of $d_f=1.570$.
Clearly, this is only a relatively weak numerical conjecture, but we nonetheless hope this to be suggestive for further analytical considerations.
\par
For the NNIM one can make a similar (albeit less obvious) assumption that $\theta=2/9=0.22\overline{2}$ which is very close to the numerically often observed value of $\theta \approx 0.22$.
Thus the value of $\theta$ in the NNIM and the LRIM with $\sigma \rightarrow \infty$ differ by factor $3/2$, which can be understood by the ratios of the growth exponents $(3/4)/(1/2)=3/2$ of those two models via Eq.~\eqref{relation_df_complete}.
Inserting $\theta=2/9$ and $\alpha=1/2$ in Eq.~\eqref{relation_df_complete} predicts $d_f=14/9=1.55\overline{5}$, in perfect agreement with the limiting value of the LRIM for $\sigma \rightarrow \infty$.
\par
Thus it is not surprising that we do not find $\theta$ for $\sigma=8$ to be compatible with the value for the NNIM.
The most robust exponent in this context appears to be the fractal dimension $d_f$, so that an universal fractal dimension for short-range-like interacting models can be conjectured.
As suggested by Eq.~\eqref{scaling} a $d_f$ compatible to the NNIM, but with a different growth exponent $\alpha$, implies a change of the persistence exponent $\theta$.
To conclude, we conjecture that even for $\sigma \rightarrow \infty$ (but still finite) as expected from $d=1$ \cite{corberi2019one}, the NNIM results are not fully recovered for all exponents.

\subsection{Small $\sigma$}
\label{LR}
Next, we target on stronger long-range interacting systems in order to check the $\sigma$ dependence of $\alpha$ and determine the values of $\theta$ and $d_f$ (and their relation).
We focus on the most long-range case of $\sigma=0.6$ we can still treat without encountering too strong finite-size effects.
Here, it is necessary to use even larger lattices with $L=4096$ to avoid finite-size effects.
Even smaller values of $\sigma$ naturally lead to significantly more pronounced finite-size effects \cite{christiansen2019non}, so that even bigger system sizes than $L=4096$ would be needed.
Since the complexity of a sweep is $V^2$ (where $V=L^2$ is the total number of spins), the computing time would increase by a factor of $16$ per sweep when doubling the system size.
Additionally, control of the onset of finite-size effects would require another factor of $2$--$3$ more computing time since on the larger lattice they would only become clearly visible at a later time.
This is currently unfeasible, so that our smallest $\sigma$ considered is $0.6$.
\par
As for $\sigma=8$ we start by investigating the snapshots of the direct and persistent lattice in Fig.~\ref{SnapSig06}.
Again, the domains grow with time for the direct lattice, although the growth appears to be faster than for $\sigma=8$.
Whether this is reflected in the growth exponent $\alpha$ or the amplitude is \emph{a priori} not clear.
For the persistent lattice, it appears that the persistent spins are more correlated, suggesting a larger fractal dimension.
Also here, the dynamics appears to be faster.
\par
We, however, first need to establish that we are dealing with a scaling phenomenon.
While especially the scaling plot of the correlation function of the direct lattice $C(r,t)$ in Fig.~\ref{ScalingSig06} is not as good as in Fig.~\ref{directCorFig}, akin to what we observed already for quenches to $T\neq 0$ ($<T_c$) \cite{christiansen2018}, we still observe satisfactory scaling for small $r$.
The (unscaled) correlation functions of the persistent lattice $D(r,t)$ plotted in Fig.~\ref{ScalingSig06Drt}(a) for several $t$ confirm again the $t$ independence for $r \ll \ell_p(t)$.
Figure~\ref{ScalingSig06Drt}(b) shows the scaling plot of $D(r,t)$ according to Eq.~\eqref{DrtScalingPlot}.
Here the data collapse looks better than for $C(r,t)$, albeit this is plotted on a log-log scale.
An objective measure of data collapse is very hard to obtain, thus we abstain ourselves from such an approach.
Note that the form of the persistent correlation function $D(r,t)$ changed (the minimum is less pronounced).
\par
The solid line in Fig.~\ref{ScalingSig06Drt}(b) is a power law  [cf.\ Eq.~\eqref{scalingDrt}] fitted to the data for $t=800$ in the range $0.03<r/\ell_p(t)<0.5$, giving $\kappa=d-d_f = 0.268(2)$ or $d_f=1.732(2)$.
This value of $d_f$ is significantly different from $d_f\approx 1.57$ found for the NNIM or the LRIM with $\sigma=8$.
Already the snapshots of the persistent lattice in Fig.~\ref{SnapSig06} indicated a larger $d_f$, so that this does not come as a surprise.
\par

\begin{table*}
  \caption{
    Values of the fractal dimension $d_f$ from power-law fits of form $f(x)=ax^{d_f-d}$ to $D(r,t)$ for different $\sigma$ and $L=2048$ for all $\sigma$.
    We fix the upper bound of the fitting range to $r_{\mathrm{max}}/\ell_p(t)=0.5$ and vary the lower bound $r_{\mathrm{min}}/\ell_p(t)$, where the corresponding values are noted in the table.
    The time used for the fit was varied from $t=800$ to $t=1200$ to allow for the longest, but still finite-size unaffected, regime.
    Mentioned are also the reduced chi-squares $\chi^2_r$, indicating the goodness of fit.
    To be able to quantitatively compare the resulting values for $\theta$ using Eq.~\eqref{relation_df_complete} (assuming $\alpha=0.75(3)$ \cite{Note1}) with the values obtained from a fit of form $P(t)=At^{-\theta_f}$, we have included both in the table.
    For the fit of $\theta_f$, the corresponding fitting ranges are also given.
  }
    \label{tab:df}
  \begin{tabular}{c|c|c|c|c|c|c|c|c|c}
    \hline
    \hline
    $\sigma$ & $0.60$ & $0.80$ & $1.00$ & $1.50$ & $1.75$ & $2.00$ & $2.25$ & $4.00$ & $8.00$\\
    \hline
    $d_f$ & $1.732(1)$ & $1.692(2)$ & $1.662(2)$ & $1.618(2)$ & $1.599(3)$ & $1.592(3)$ & $1.587(3)$ & $1.571(2)$ & $1.570(2)$ \\
    \hline
    $t$ & $800$ & $800$ & $800$ & $800$ & $1200$ & $1200$ & $1200$ & $1200$ & $1200$ \\
    \hline
    $r_{\mathrm{min}}/\ell_p(t)$ & $0.03$ & $0.05$ & $0.1$ & $0.15$ & $0.15$ & $0.15$ & $0.15$ & $0.15$ & $0.15$ \\
    \hline
    $\chi^2_r$ & $6(2)$ & $5(2)$ & $1.4(5)$ & $1.1(3)$ & $3(2)$ & $3(1)$ & $3(2)$ & $1.1(7)$ & $1.9(9)$ \\
    \hline
    $\theta=(d-d_f)\alpha$ & $0.201(8)$ & $0.231(9)$ & $0.25(1)$ &  $0.29(1)$ & $0.30(1)$ & $0.31(1)$ & $0.31(1)$ & $0.32(1)$ & $0.32(1)$\\
    \hline
    $\theta_f$ & $0.1998(5)$ & $0.2315(9)$ & $0.258(1)$ & $0.288(1)$ & $0.288(1)$ & $0.302(1)$ & $0.299(1)$ & $0.317(1)$ & $0.321(2)$ \\
    \hline
    Fitting range of $\theta_f$ & $[20,200]$ & $[30,400]$ & $[40,700]$ & $[50,900]$ & $[50,900]$ & $[50,950]$ & $[50,950]$ & $[50,950]$ & $[100,1000]$ \\
    \hline
    \hline
  \end{tabular}
\end{table*}
We next consider the direct length $\ell(t)$ plotted for the three different system sizes in Fig.~\ref{length_sig_06}(a).
The dashed black line is a power law with exponent $\alpha=3/4$, which appears consistent with our data.
A direct fit of the data for $L=4096$ in the range from $100 \le t \le 750$ provides $\alpha=0.723(8)$.
The finite-size effects become apparent from the earlier onset of a downward tendency of the data for smaller values of $L$.
Note that the finite-size effects play a role even before they are visible in this figure.
The dashed lines in the same color as the solid lines show $\ell(t)/t^{3/4}$ to demonstrate that asymptotically the data is consistent with this exponent, as the region with a constant value increases for increasing system size.
\par
\begin{figure}
\includegraphics{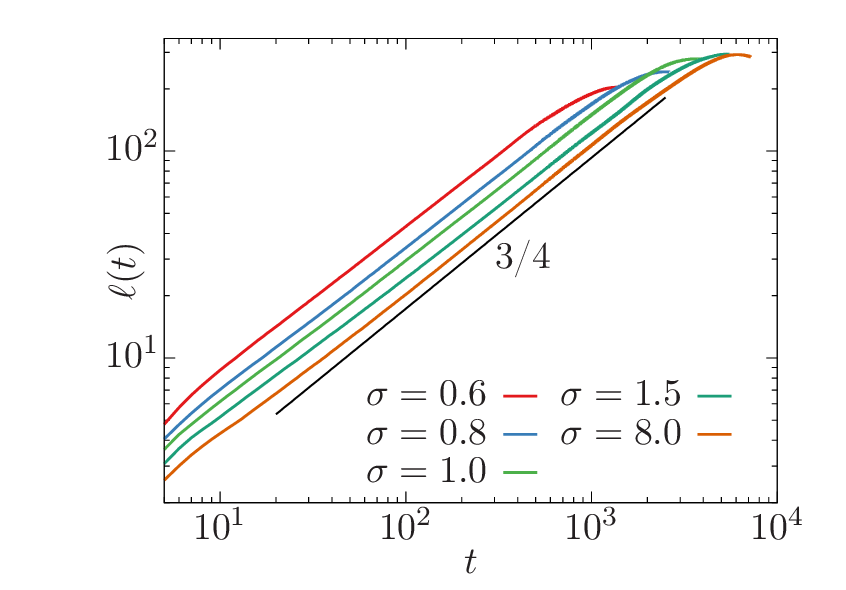}
 \caption{Direct length $\ell(t)$ for different values of $\sigma \in [0.6,8]$ with system size $L=2048$. The solid black line corresponds to a power law $\propto t^{\alpha}$, with growth exponent $\alpha=3/4$ as estimated before.}
 \label{FigLdiffS}
\end{figure}
\begin{figure}
  \includegraphics{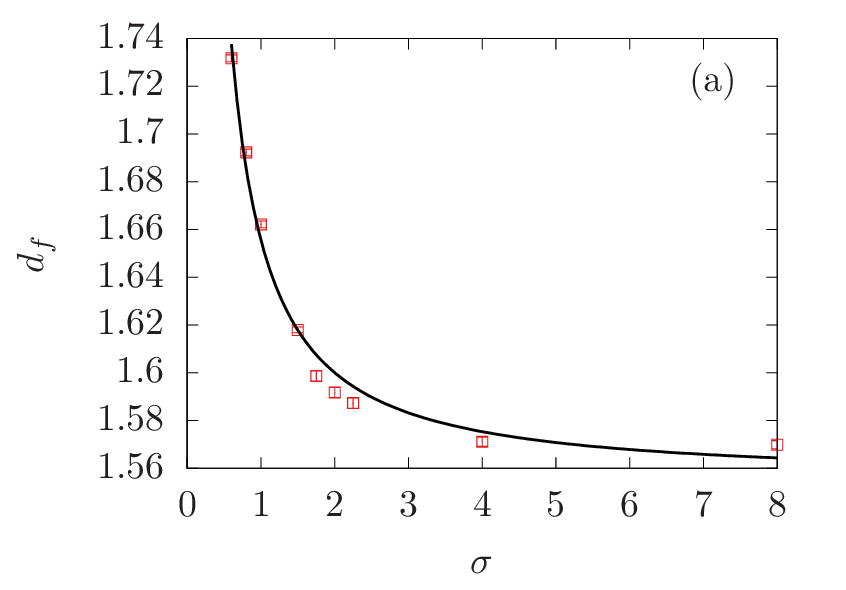}
  \includegraphics{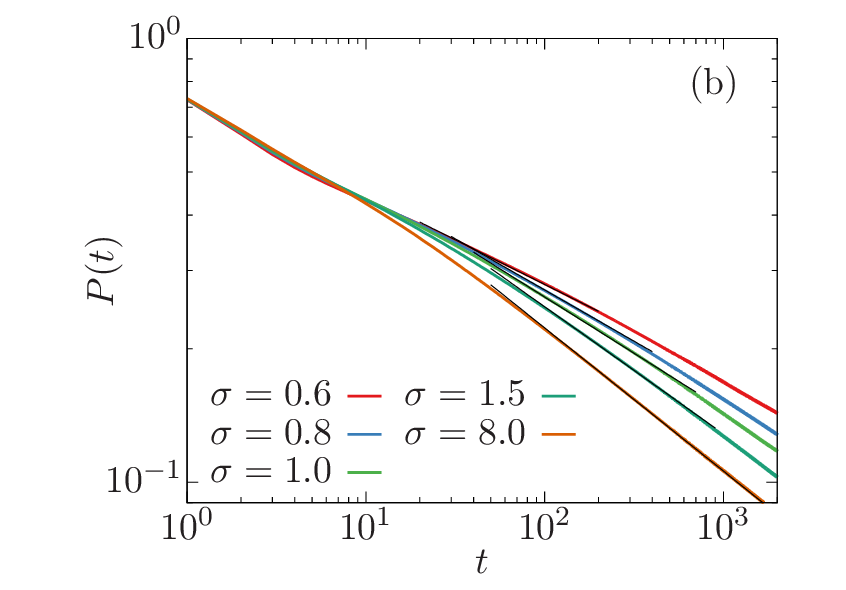}
  \caption{
    (a) Fractal dimension $d_f$ against $\sigma$ as compiled in Table~\ref{tab:df}.
    The solid black line is a fit of form $d_f(\sigma)=d_{f,\infty}+A\sigma^{-B}$ with $d_{f,\infty}=1.55\overline{5}$ fixed and fit parameters $A=0.1001(7)$ and $B=1.17(2)$.
    (b) Persistence probability $P(t)$ for different $\sigma$ versus time $t$.
    The perfectly matching solid black lines are power laws $t^{-\theta}$ with exponents $\theta$ obtained from Eq.~\eqref{relation_df_complete} by plugging in the values of $d_f$ from Table~\ref{tab:df} and using $\alpha=3/4$.
  }
  \label{FigPtdiffL}
\end{figure}
The finite-size effects for $\ell_p(t)$ in Fig.~\ref{length_sig_06}(b) are much less pronounced and only by a very careful investigation occur with slightly \emph{higher} values of $\ell_p(t)$ for the \emph{smaller} system sizes, whereas the finite-size effects for the direct length $\ell(t)$ appear in opposite direction, i.e., towards \emph{smaller} values and not \emph{higher} values.
Performing a fit in the range $100 \le t \le 750$ for $\ell_p$ gives $\alpha=0.773(5)$.
Here, also, we instead plot as a dashed black line a power law with exponent $\alpha=3/4$ and the dashed lines in the same color as the original data are $\ell_p(t)/t^{3/4}$.
There exists a region where $\ell_p(t)/t^{3/4}$ is constant.
Thus we conclude that also in this case, the growth exponent is approximately given by $\alpha \approx 3/4$, suggesting that $\alpha$ is $\sigma$ independent.
\par
In Fig.~\ref{PersistenceSig06} we show the persistence probability for $\sigma=0.6$ and different $L$.
The dashed black line is a power law $\sim t^{-\theta}$, where $\theta =  0.1964(6)$ was obtained from a fit to the data for $L = 4096$ in the range $20 \le t \le 200$.
To consolidate this, as before we plot $P(t)/t^{-\theta}$ as dashed line which is approximately constant over the entire time frame and thus confirms the fit.
By using the scaling relation~\eqref{scaling} with $\kappa = 0.268(2)$ and $\alpha = 0.75(3)$ \footnote{The error bar on $\alpha$ reflects the systematic deviations stemming from the fits of $\ell(t)$ and $\ell_p(t)$ using different $\sigma$ and fitting ranges.} one gets $\theta =  0.201(8)$ which would also be compatible with the data from roughly $t \approx 30$ to $t \approx 500$.
Using $\alpha$ estimated from the growth of $\ell_p$ changes this estimate only slightly to $\theta = 0.207(2)$. 

\subsection{Intermediate $\sigma$}
\label{diffSig}
We now inspect the scaling behavior for $\sigma$ in between the two extremes studied before, covering a wide range of interactions. 
In Fig.~\ref{FigLdiffS} we plot the direct length $\ell(t)$ on a log-log plot for $\sigma=0.6$, $0.8$, $1.0$, $1.5$, and $8$, where we use $L=2048$ for all $\sigma$ in order to allow for an easier direct comparison.
It is apparent that all data are more or less parallel to each other in a relatively long region and thus the estimate of $\alpha=3/4$ for all $\sigma$ is compatible with the data.

In Table~\ref{tab:df} we present results from power-law fits of form~\eqref{scalingDrt} to $D(r,t)$, giving estimates for $d_f$ and the goodness of fit $\chi^2_r$.
We perform the fits in the range $r_{\mathrm{min}}/\ell_p(t)$ to $r_{\mathrm{max}}/\ell_p(t)$, where we vary $r_{\mathrm{min}}/\ell_p(t)$ and set $r_{\mathrm{max}}/\ell_p(t)=0.5$ for all $\sigma$.
We also give the time $t$ used, which we want to choose as big as possible without experiencing finite-size effects, so that we chose $t=800$ for $\sigma\leq 1.5$ and $t=1200$ otherwise.
The reduced chi-square value $\chi^2_r$ is relatively big for $\sigma=0.6$ and $\sigma=0.8$ (also due to the smaller system size $L=2048$), but since we have carefully estimated the finite-size effects in the previous section these fits should nonetheless be appropriate.
Additionally, we present the values of $\theta_f$ obtained from fits of the form $P(t)=At^{-\theta_f}$ and the corresponding fitting range.
This allows to directly test the validity (and thereby general applicability) of Eq.~\eqref{relation_df_complete} by comparing $\theta=(d-d_f)\alpha$ with $\theta_f$, which agree very well.

To get a better idea of the functional dependency obtained for the exponents, we plot $d_f$ versus $\sigma$ in Fig.~\ref{FigPtdiffL}(a).
From our results for quenches to $T\neq 0$, one would expect some kind of transition at $\sigma=1$, whereas from equilibrium studies one could expect a transition at $\sigma=2$ \cite{fisher1972critical} or $\sigma=1.75$ \cite{sak1973recursion}.
The correct value for the crossover from the indermediate to short-range-like regime in equilibrium is still disputed in the literature \cite{luijten2002boundary,picco2012critical,blanchard2013influence,angelini2014relations,defenu2015fixed,horita2017upper}.
In our setting, however, no distinct crossover at any $\sigma$ is observed.
Thus one has to conclude, that this equilibrium phase transition does not manifest itself in the nonequilibrium dynamical behavior at $T=0$.
We rather observe a smooth approach of $d_f$ to a value compatible with the NNIM fractal dimension.
We are not aware of any theoretical conjecture for the functional dependency of $d_f$ on $\sigma$.
We therefore empirically fitted a power law of the form $d_f(\sigma)=d_{f,\infty}+A\sigma^{-B}$, where $d_{f,\infty}=1.55\overline{5}$ is the estimate of $d_f$ obtained by assuming $\alpha=3/4$ and $\theta=1/3$.
The corresponding fit is also shown in Fig.~\ref{FigPtdiffL}(a), having $A=0.1001(7)$ and $B=1.17(2)$.
This gives us a rough idea about the functional dependency, however, can not be taken too seriously, as the fit has $\chi^2_r \approx 17$ which of course indicates a very bad fit.
\par
We want to point out that the persistence probability is often investigated as a function of the characteristic length $\ell(t)$ as already mentioned in the previous sections, i.e., $P(t) \sim \ell(t)^{-\overline{\theta}}$, where if $\ell(t) \sim t^{\alpha}$ one has $\overline{\theta}=\theta/\alpha=d-d_f$ (but $P(\ell(t))$ may also be investigated in situations, where $\ell(t)$ is not a power law).
Thus, instead of considering $d_f$, we also could have investigated $\overline{\theta}$ and obtained that $\overline{\theta}$ would approach the same value $\theta/\alpha=0.44\overline{4}$ for all short-range-like models (by assuming $\theta=2/9$ and $\alpha=1/2$ for the NNIM and $\theta=1/3$ and $\alpha=3/4$ for the LRIM with $\sigma \rightarrow \infty$).
In this sense, none of the observed values appear ``odd'' and their relationship can be well understood.
\par
In Fig.~\ref{FigPtdiffL}(b) we show the persistence probability $P(t)$ for different $\sigma$.
We use Eq.~\eqref{relation_df_complete} to obtain estimates for $\theta$, which are used in the power laws $t^{-\theta}$ plotted as solid black lines.
For a relatively large range, the power laws are consistent with the data for $P(t)$.
Of course, the scaling regime gets smaller the smaller $\sigma$ is.
Having considered the finite-size effects carefully, however, we are confident that the values for the exponents we quote are the true asymptotic values for all $\sigma$.

\section{Conclusion}
\label{conclusion}
We have studied the zero-temperature coarsening of the two-dimensional long-range Ising model with non-conserved order parameter by tuning the degree of the long-range interactions via the power-law exponent $\sigma$.
It is found that the growth exponent $\alpha \approx 3/4$ is independent of $\sigma$ and in the limit $\sigma \rightarrow \infty$ does not seem to approach the value $\alpha=1/2$ of the nearest-neighbor model. 
For our most short-range-like case of $\sigma=8$, we find that the fractal dimension is compatible with the value found for the nearest-neighbor Ising model and reads $d_f \approx 1.57$.
Evidence was provided in favor of the relation $d-d_f=\theta/\alpha$, which relates the nonequilibrium exponents.
Here $\theta$ is the persistence exponent and $d$ is the spatial dimension.
For $\sigma=8$ we find $\theta \approx 0.32$, which \emph{a priori} is significantly different from the value for the nearest-neighbor Ising model with $\theta \approx 0.22$.
However, this can be still understood, since those two exponents just differ by a factor of $\approx 1.5$, which is exactly the ratio of the growth exponents for those two models as expected from the above relationship.
In fact, if one considers the scaling of the persistence probability $P(t)$ with the characteristic length scale $\ell(t)$, $P(t) \sim \ell(t)^{-\overline{\theta}}$, as in Refs.~\cite{ispolatov1999persistence,bray1994non}, $\overline{\theta}=\theta/\alpha$ of the long-range model would agree in the asymptotic limit $\sigma \rightarrow \infty$ with $\overline{\theta}$ of the nearest-neighbor model.
\par
In the most long-range-like system under consideration with $\sigma=0.6$, we find that above relation relating the nonequilibrium exponents still is valid.
Here, we find $d_f\approx 1.73$ and thus $\theta \approx 0.20$.
The value of $\theta$ is most probably only coincidentally close to the nearest-neighbor Ising model value.
\par
Finally, when investigating a range of different $\sigma$ one finds that $d_f$ (and thereby $\theta$) varies continuously with $\sigma$.
There does not appear to be any distinct crossover.
\par
As a further direction of investigation, one could redo this kind of analysis also in $d=1$ and $d=3$ dimensions.
Such an endeavor could be even helpful in (more accurate) estimates of the fractal dimension exponent found in the nearest-neighbor Ising model in $d=3$ at zero temperature, where finite-size effects are enormous, since much less simulations get trapped in (meta-) stable confirmations and the ground state is reached much more often.
\par
After completion of this work, we became aware of the very recent preprint \cite{agrawal2020kinetics} where the authors focus on the growth exponent $\alpha$ and also find $\ell(t) \sim t^{3/4}$ independent of $\sigma$.
They additionally provide some arguments for the value of this exponent using a simplified model, but there does not appear to be a \emph{simple} explanation for the observed rational exponent.

\begin{acknowledgments}
This project was funded by the Deutsche Forschungsgemeinschaft (DFG, German Research Foundation) under project No. 189\,853\,844 -- SFB/TRR 102 (project B04), and the Deutsch-Französische Hochschule (DFH-UFA) through the Doctoral College ``$\mathbb{L}^4$'' under Grant No.\ CDFA-02-07. We further acknowledge support by the Leipzig Graduate School of Natural Sciences ``BuildMoNa''.
\end{acknowledgments}

 \end{document}